\documentclass[a4paper,11pt]{article}

\usepackage{jcappub} 
\usepackage[T1]{fontenc} 
\usepackage[utf8]{inputenc}
\usepackage{nicefrac}
\usepackage{amsfonts}
\usepackage{lmodern}
\usepackage{amsthm,graphicx}
\usepackage{epsfig}
\usepackage[normalem]{ulem}
\usepackage{latexsym,amssymb} 
\usepackage{amsmath,mathtools}
\usepackage{xcolor,xspace}
\usepackage{siunitx}
\usepackage{multirow}
\usepackage[font=footnotesize, labelfont=bf]{subcaption}
\usepackage{booktabs} 
\usepackage{empheq}

\newcommand{\Planck}{\textit{Planck}\ }
\usepackage{color}

\notoc

\title{A Minkowski Functional Analysis of the Cosmic Microwave Background Weak Lensing Convergence}

\author{Jan Hamann and Yuqi Kang}
\affiliation{Sydney Consortium for Particle Physics and Cosmology, School of Physics, The University
of New South Wales, Sydney NSW 2052, Australia}

\emailAdd{jan.hamann@unsw.edu.au}
\emailAdd{yuqi.kang@unsw.edu.au}

\abstract{Minkowski functionals are summary statistics that capture the geometric and morphological properties of fields. They are sensitive to all higher order correlations of the fields and can be used to complement more conventional statistics, such as the power spectrum of the field.  We develop a Minkowski functional-based approach for a full likelihood analysis of mildly non-Gaussian sky maps with partial sky coverage.  Applying this to the inference of cosmological parameters from the \textit{Planck} mission's map of the Cosmic Microwave Background's lensing convergence, we find an excellent agreement with results from the power spectrum-based lensing likelihood.  While the non-Gaussianity of current CMB lensing maps is dominated by reconstruction noise, a Minkowski functional-based analysis may be able to extract cosmological information from the non-Gaussianity of future lensing maps and thus go beyond what is accessible with a power spectrum-based analysis.  We make the numerical code for the calculation of a map's Minkowski functionals, skewness and kurtosis parameters available for download from \href{https://github.com/Kang-Yuqi/MF_lensing}{GitHub}.\footnote{\href{https://github.com/Kang-Yuqi/MF_lensing}{\texttt{https://github.com/Kang-Yuqi/MF\_lensing}}}}

\begin{document}
\begin{flushright}
\large \tt{CPPC-2023-10}
\end{flushright}

\maketitle
\flushbottom

\newpage
\tableofcontents

\section{Introduction}

    Over the past couple of decades, our knowledge of the Universe has greatly advanced, thanks to increasingly precise measurements of cosmological observables.  Of particular importance here are the probes of cosmic structure, such as the anisotropies of the Cosmic Microwave Background (CMB)~\cite{Planck:2018nkj}, or the inhomogeneities of the matter density in the local Universe, probed for instance via the distribution of galaxies~\cite{Gil-Marin:2014sta,DES:2021wwk} or weak gravitational lensing effects~\cite{Bartelmann:1999yn,Lewis:2006fu,Planck:2018lbu}.

    For the most part, on sufficiently large scales or at sufficiently early times, the relative perturbations are small, their dynamics well-approximated by linear perturbation theory, and they can be modelled as statistically isotropic Gaussian random fields.  In this case, the power spectrum of the perturbations is an ideal summary statistic and has therefore become the standard tool for likelihood-based inference of cosmological parameters.  
    However, at later times and on small enough scales, non-linear physics will lead to the breakdown of linear perturbation theory and induce deviations from Gaussianity – these may contain valuable cosmological information to which a power spectrum-based analysis would be blind.  
    
    With the increasing observational precision to be expected from upcoming cosmological missions~\cite{Abazajian:2019eic,LSSTScience:2009jmu,Amendola:2016saw}, we will also become more sensitive to these non-Gaussian signatures, so it is well worth contemplating alternative analysis methods that may complement the power spectrum in this regard.  One idea is to eschew the idea of a likelihood altogether and use machine-learning techniques to perform likelihood-free inference~\cite{Alsing:2019xrx, Jeffrey:2020xve}.  Alternatively, one could stick with a likelihood function, but consider different summary statistics.  A full systematic approach to this problem in terms of higher order correlators (via the bispectrum, trispectrum, etc.) can quickly become computationally prohibitive, and it may therefore be useful to also consider simpler statistics, such as peak counts~\cite{Liu:2014fzc}, the one-point probability distribution function~\cite{Liu:2016nfs} or Minkowski functionals (MFs)~\cite{Mink-1903} – which we will study in the present work.  
    
    Minkowski functionals are statistics that describe morphological properties of fields.  They have previously been applied to, e.g., the search for non-Gaussianities or statistical anisotropies in CMB temperature maps from the WMAP~\cite{Eriksen:2004df, Hikage:2006fe} and  \Planck surveys~\cite{Buchert:2017uup, Planck:2019evm}, CMB polarisation maps~\cite{Carones:2022rbv,CarronDuque:2023bph}, the density field from the SDSS~\cite{Hikage:2003fc} and SDSS-III \cite{Appleby:2021xoz} data releases, the analysis of the WISExSuperCOSMOS galaxy field~\cite{Novaes:2018kbw}, CFHTLenS lensing data~\cite{Petri:2015ura} and for forecasts for future galaxy and galaxy lensing surveys~\cite{Marques:2018ctl, Parroni:2019snd, Zurcher:2020dvu,Liu:2022vtr}.

    Our goal in this article is to develop a MF-based likelihood approach to parameter inference and as a proof of principle apply it to the \Planck CMB weak gravitational lensing convergence map.  There are three reasons why one would expect a map of the CMB lensing convergence to deviate from Gaussianity: firstly, due to gravitational lensing being an intrinsically non-linear process (this is certainly true for a single ``lensing event'' though one could make an argument based on the central limit theorem that the non-Gaussianity will be washed out by the fact that the CMB photons are subject to many lensing events on their way from the last scattering surface).  Secondly, because the lens itself (i.e., the large scale structure at redshifts $z \sim \mathcal{O}(1)$) is becoming non-Gaussian at late times due to the non-linear growth of structure, and thirdly, due to non-Gaussianity introduced by the lensing reconstruction.  As it turns out, in the \Planck map the latter effect is dominant, and thus we will not be able to extract any cosmological information from its higher order correlations, but this situation may change in future observations.  In any case, at the very least our analysis  provides an independent consistency check of the standard power spectrum-based likelihood.

    The structure of the paper is as follows: in section~\ref{sec:Planck}, we briefly talk about the CMB lensing map reconstructed from \Planck observations. In section~\ref{sec:Mink}, we review some basic properties Minkowski Functionals, discuss how to estimate them from full-sky pixelated maps and how to deal with incomplete sky coverage. We introduce and validate a MF-based likelihood for the purpose of constraining cosmological parameters in Section~\ref{sec:likelihood} and apply it to \Planck lensing data in Section~\ref{sec:results}.  Finally, we present our conclusions in Section~\ref{sec:conclusions}.

\section{The \Planck CMB lensing map} \label{sec:Planck}

    While the formalism we develop in this paper is quite general and can in principle be applied to arbitrary fields on a partially covered sphere, the main motivation for our endeavours is an application to CMB lensing data from the \Planck 2018 data release~\cite{Planck:2018lbu}.  The CMB lensing effect itself is of course not directly observable, but it can be reconstructed from CMB temperature and polarisation maps, using for instance quadratic estimators~\cite{Carron:2017mqf}.  This approach has been used on \Planck temperature and polarisation data, resulting in a map of the lensing potential $\psi$~\cite{Planck:2018lbu}, or alternatively the lensing convergence $\kappa$ on which we will focus our analysis here – and which is related to the lensing potential via 
    \begin{equation}
        \kappa = -\dfrac{1}{2}\nabla^{2} \psi.
    \end{equation} 

\begin{figure}[ht]
		\centering
		\includegraphics[width=0.7\textwidth]{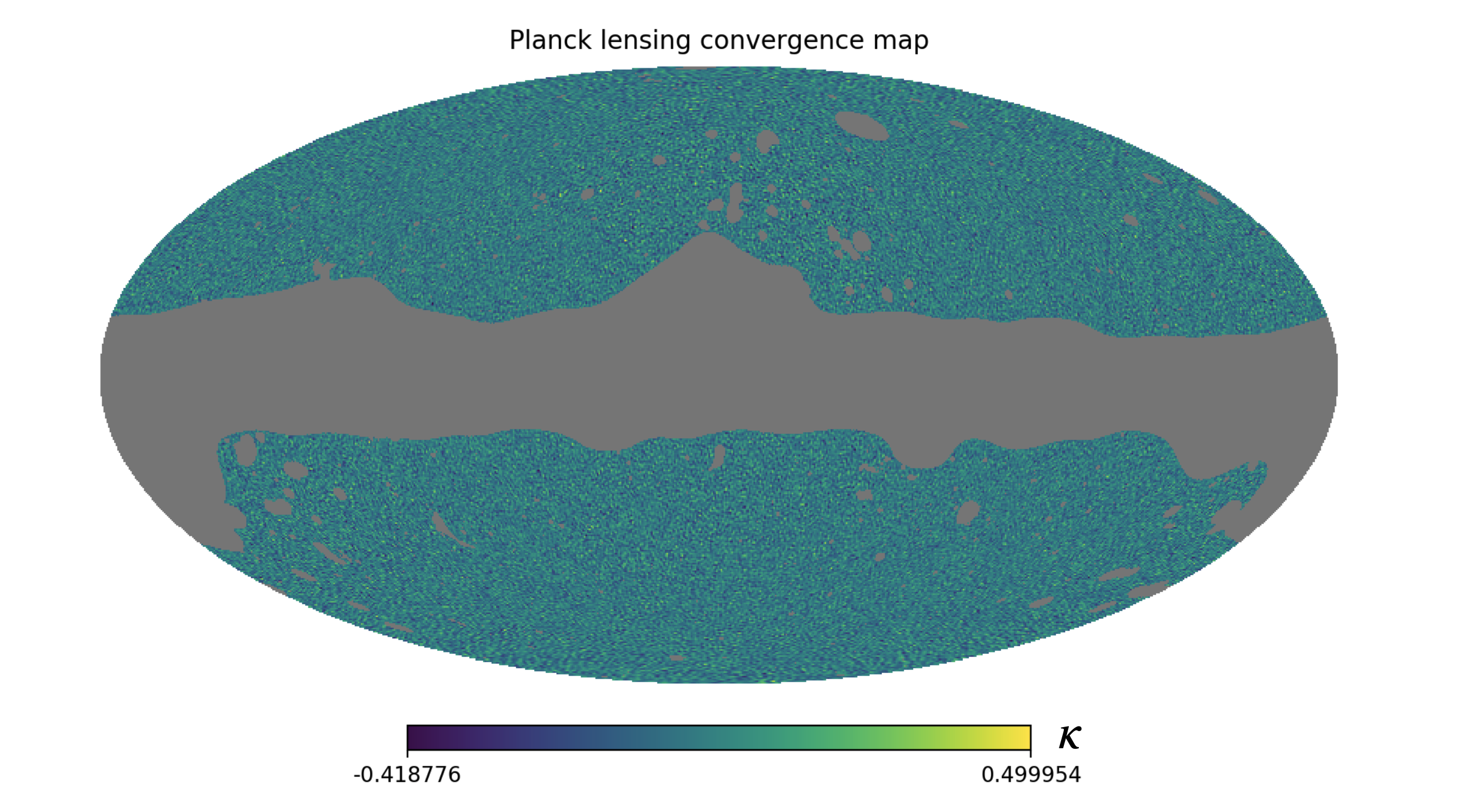}
		\caption{\Planck 2018 map of the lensing convergence $\kappa$, with the \Planck lensing reconstruction mask overlaid in grey.}
		\label{fig:Planck}
		\end{figure}

    We show the \Planck 2018 lensing convergence map in Figure~\ref{fig:Planck}, along with the \Planck lensing reconstruction mask.  The mask removes foreground-dominated parts of the sky such as the galactic plane, as well as point sources, and retains a sky fraction of $f_\mathrm{sky} \simeq 0.671$.  We note that this map is in fact dominated by reconstruction noise at most scales; the signal-to-noise reaches a maximum of about 1 on angular scales around 5 degrees (corresponding to multipoles $\ell \sim 40$).  Additionally, as a result of \textit{Planck}'s scanning pattern, the noise in temperature and polarisation maps has a substantial anisotropy which leads to anisotropic noise in the reconstructed CMB lensing map as well.
    
    For consistency with the \Planck power spectrum-based analysis of Ref.~\cite{Planck:2018lbu}, we discard information on the largest and smallest scales, and restrict our analysis to the conservative range $8 \leq \ell \leq 400$.

	Besides the \Planck lensing map, we shall also make use of the set of 300 \Planck Full Focal Plane (FFP10) simulation lensing maps~\cite{Planck:2018lkk} to construct realistic covariance matrices and correct biases associated with anisotropies and non-Gaussianity introduced by the reconstruction process, see Section~\ref{sec:Planck_noise}.

\section{Minkowski Functionals \label{sec:Mink}}

	Given a real scalar function $f(\bf{x})$ on a 2-dimensional surface, we can define an excursion set $\Sigma_\nu$ with a threshold $\nu$, where $\Sigma_\nu \equiv \{ \mathbf{x} : f( \mathbf{x})>\nu\}$. The MFs encode the morphological properties of these excursion set. In two dimensions, there are three MFs, namely the area ($V_{0}$), circumference ($V_{1}$) and Euler characteristic ($V_{2}$):
	
	\begin{equation}\label{MF0}
		V_{0}(\nu) = \dfrac{1}{A}\int_{\Sigma_{\nu}} \mathrm{d}a,
	\end{equation}
	
	\begin{equation}\label{MF1}
		V_{1}(\nu) = \dfrac{1}{4A}\int_{\partial \Sigma_{\nu}} \mathrm{d}l,
	\end{equation}		

	\begin{equation}\label{MF2}
		V_{2}(\nu) = \dfrac{1}{2\pi A}\int_{\partial \Sigma_{\nu}} \mathcal{K} \, \mathrm{d}l,
	\end{equation}		
	where $A$ is the area of the field's support, $\mathrm{d}a$ and $\mathrm{d}l$ are the surface and line elements, respectively, $\partial \Sigma_{\nu}$ is the boundary of $\Sigma_{\nu}$ and $\mathcal{K}$ is the geodesic curvature of $\partial \Sigma_{\nu}$. The Euler characteristic can also be understood as the difference between the number of connected regions above the threshold $\nu$ and the number of connected regions below $\nu$.

	MFs contain information of all higher order correlations in the field, which makes them useful tools to analyse non-Gaussian fields. Moreover, the third functional $V_{2}$ is invariant under diffeomorphisms of the field, and thus very robust with respect to observational systematics.

\subsection{Theoretical prediction of MFs}
	If one wants to use MFs for a quantitative cosmological data analysis, it is crucial to have accurate theoretical predictions. As mentioned above, MFs are sensitive to the Gaussian \emph{and} non-Gaussian information contained in the observed field. 
 \subsubsection{Gaussian random fields \label{sec:grf}}
    For a Gaussian random field, the information is completely contained in the power spectrum, which can be calculated with a Boltzmann solver, e.g., \texttt{CLASS}~\cite{Blas:2011rf} or \texttt{CAMB}~\cite{Lewis:1999bs}.\footnote{Predicting the complete non-Gaussian information of the full-sky CMB lensing potential on the other hand is a much more challenging task and would necessitate numerically expensive $N$-body simulations, (e.g.~\cite{Carbone:2007yy}).  
	For the \Planck lensing map, however, the non-Gaussianity is completely dominated by reconstruction noise~\cite{Liu:2016nfs}, which is independent of the cosmological model.  As described in more detail in Sec.~\ref{sec:Planck_noise}, we can thus use the FFP10 simulations to model the non-Gaussian contribution.}

	The MFs of a Gaussian random field can thus be calculated analytically via its power spectrum, and the corresponding expressions for MFs in Euclidean space of arbitrary dimension were first derived by Adler~\cite{Adler:1981xxx} (see also earlier work by Doroshkevich~\cite{Doroshkevich:1970xyz} which was limited to the Euler characteristic, and the later work of Tomita~\cite{tomita:1986}).  In the specific case of a Gaussian field in the Euclidean plane, we have~\cite{Schmalzing:1997uc}:
		
	\begin{equation}\label{V0_PS}
	   V_{0}(\nu) = \dfrac{1}{2}-\dfrac{1}{2}\, \mathrm{erf} \left( \dfrac{\nu-\mu}{\sqrt{2}\sigma_{0}}\right),
	\end{equation}
		
	\begin{equation}\label{V1_PS}
		V_{1}(\nu) = \dfrac{\sigma_{1}}{8\sigma_{0}} \, \exp\left( -\dfrac{\left( \nu-\mu\right)^{2} }{2\sigma_{0}^{2}}\right) ,
	\end{equation}
		
	\begin{equation}\label{V2_PS}
		V_{2}(\nu) = \dfrac{1}{\left(2\pi\right)^{\frac{3}{2}}}\dfrac{\sigma_{1}^{2}}{\sigma_{0}^{2}} \, \left( \dfrac{\nu-\mu}{\sqrt{2}\sigma_{0}}\right) \, \exp\left( -\dfrac{\left( \nu-\mu\right)^{2} }{2\sigma_{0}^{2}}\right),
	\end{equation}
	where $\mu$ is the average value of the field and $\sigma_{0}$ and $\sigma_{1}$ are the standard deviation and first moment of the field, respectively.\footnote{Since the Euler characteristic of a sphere is known to be 2, one must in principle modify Equation~\ref{V2_PS} when generalising these expressions to spherical geometry in order to ensure that one recovers the correct result in the limit where the excursion set covers the entire sphere ($\nu \rightarrow -\infty$).  The correction term $\delta V_2(\nu)$ can be computed using the Gaussian Kinematic Formula~\cite{Adler&Taylor} and is given by~\cite{Fantaye:2014vua}
    \begin{equation}\nonumber
        \delta V_2(\nu) = 2 \left( 1 - \mathrm{erfc} \left(\frac{\nu -\mu}{\sigma_0}\right)\right).
    \end{equation}
    However, as argued in Section~2.2 of Ref.~\cite{Schmalzing:1997uc}, instead of the Euler characteristic one may just as well use the integrated curvature of the field along the boundary of the excursion set as a morphological descriptor (the two coincide in Euclidean space).  In the following, we will stick to the latter and continue to use Equation~\ref{V2_PS}, since the curvature integral is more straightforwardly evaluated numerically on a given map (see Equation~\ref{V2_map1}).}  
 For a Gaussian random field on a sphere,  $\sigma_{0}$ and $\sigma_{1}$ are directly related to the angular power spectrum of the field $\mathcal{C}_\ell$ via
		
	 \begin{equation}\label{sigma}
		 \sigma_{0}^{2} = \frac{1}{4\pi}\sum_{\ell}\left(2\ell+1\right) \, \mathcal{C}_\ell,
	 \end{equation}
		 
	 \begin{equation}\label{tau}
		 \sigma_{1}^{2} = \dfrac{1}{4\pi}\sum_{\ell}\left(2\ell+1\right)\ell(\ell+1) \, \mathcal{C}_\ell.
	 \end{equation}

\subsubsection{Mildly non-Gaussian random fields\label{sec:nGcorr}}
	For a weakly non-Gaussian field, the MFs can be expanded in terms of the field's standard deviation $\sigma_{0}$ as 
	\begin{equation}\label{V_pert}
	V_{k}(\nu) = V_{k}^\mathrm{G}(\nu) + V_{k}^{(0)}(\nu)\sigma_{0} + V_{k}^{(1)}(\nu)\sigma_{0}^{2} + ...,
	\end{equation}
	where $V_{n}^\mathrm{G}$ indicates the Gaussian contribution to the MFs, and $V_{n}^{0}$ and $V_{n}^{1}$ are the first and second order non-Gaussian corrections. 
	The explicit expressions for the first and second order corrections were derived by Matsubara et al.~\cite{Matsubara:2003yt,Matsubara:2010te,Matsubara:2020fet}, with the first order corrections linear in the skewness parameters $S^{(n)}$ of the field and the second order corrections having contributions quadratic in the skewness parameters and linear in the kurtosis parameters $K_{m}^{(n)}$:

\begin{align}
		V_{0}(\nu)=&\frac{1}{\sqrt{2\pi}}e^{-\nu^{2}/2}\left[H_{-1}(\nu) + \frac{S^{(0)}}{6} H_{2}(\nu)\sigma_{0} +\left[ \frac{(S^{(0)})^{2}}{72} H_{5}(\nu)+\frac{K^{(0)}}{24} H_{3}(\nu)\right]\sigma_{0}^{2}\right],\\
		V_{1}(\nu)=& \frac{\sigma_{1}}{8\sqrt{2}\sigma_{0}}e^{-\nu^{2}/2}\left[H_{0}(\nu) + \left[ \frac{S^{(0)}}{6} H_{3}(\nu) + \frac{S^{(1)}}{3} H_{1}(\nu)\right]\sigma_{0} \right.\nonumber\\ 
		&+ \left[ \left( \frac{K^{(0)}}{24} + \frac{S^{(0)}S^{(1)}}{18}\right)H_{4}(\nu) + \left( \frac{K^{(1)}}{8} - \frac{(S^{(1)})^{2}}{18}\right)H_{2}(\nu) \right.\\
		&\left.\left.+ \left( \frac{K_{2}^{(2)}-K_{1}^{(2)}}{16}\right)H_{0}(\nu) + \frac{1}{72}\left(S^{(0)}\right)^{2}H_{6}(\nu) \right]\sigma_{0}^{2}\right],\nonumber\\
		V_{2}(\nu)=& \frac{1}{(2\pi)^{3/2}}\left(\frac{\sigma_{1}}{\sqrt{2}\sigma_{0}}\right)^{2}e^{-\nu^{2}/2} \left[ H_{1}(\nu) + \left[ \frac{S^{(2)}}{3}H_{0}(\nu) + \frac{2S^{(1)}}{3}H_{2}(\nu) + \frac{S^{(0)}}{6}H_{4}(\nu)\right]\sigma_{0}\right.\nonumber\\
		&+ \left[ \left( \frac{K_{2}^{(2)}}{4}- \frac{2S^{(1)}S^{(2)}}{9}\right)H_{1}(\nu) + \left(\frac{K^{(1)}}{4}+\frac{S^{(0)}S^{(2)}}{18}\right)H_{3}(\nu)\right.\\
		& + \left.\left.\left(\frac{K^{(0)}}{24}+\frac{S^{(0)}S^{(1)}}{9}\right)H_{5}(\nu) + \frac{1}{72}\left(S^{(0)}\right)^{2}H_{7}(\nu)\right]\sigma_{0}^2\right],  \nonumber  
\end{align}

	where $H_n(\nu) \equiv (-1)^n \, e^{\nu^2/2}\, \frac{\mathrm{d}^n}{\mathrm{d}\nu^n} e^{-\nu^2/2}$ are the probabilist's Hermite polynomials for $n \ge 0$ and $H_{-1}(\nu) \equiv \sqrt{\frac{\pi}{2}} \, e^{\nu^2/2} \,\mathrm{erfc}\left(\frac{\nu}{\sqrt{2}}\right)$. The skewness and kurtosis parameters $S^{(n)}$ and $K_{m}^{(n)}$ are related to higher order correlations of the field $f$ and its derivatives, as well as $\sigma_{0}$,$\sigma_{1}$, 

	\begin{equation}\label{S_para}
	\begin{aligned}
    	&S^{(0)}=\frac{\left\langle f^{3}\right\rangle_{\mathrm{c}}}{\sigma_{0}^{4}}, \quad S^{(1)}=\frac{3}{2} \frac{\left\langle f|\mathbf{\nabla} f|^{2}\right\rangle_{\mathrm{c}}}{\sigma_{0}^{2} \sigma_{1}^{2}} ,\\
    	&S^{(2)}=-3\frac{\left\langle|\nabla f|^{2} \Delta f\right\rangle_{\mathrm{c}}}{\sigma_{1}^{4}},
    	\end{aligned}
    	\end{equation}
    	\begin{equation}\label{K_para}
    	\begin{aligned}
    	&K^{(0)}=\frac{\left\langle f^{4}\right\rangle_{\mathrm{c}}}{\sigma_{0}^{6}}, \quad K^{(1)}=2 \frac{\left\langle f^{2}|\nabla f|^{2}\right\rangle_{\mathrm{c}}}{\sigma_{0}^{4} \sigma_{1}^{2}}, \\
    	&K_{1}^{(2)}=-\frac{4\left\langle f|\nabla f|^{2} \Delta f\right\rangle_{\mathrm{c}}+\left\langle|\nabla f|^{4}\right\rangle_{\mathrm{c}}}{\sigma_{0}^{2} \sigma_{1}^{4}}, \\
    	&K_{2}^{(2)}=-\frac{4\left\langle f|\nabla f|^{2} \Delta f\right\rangle_{\mathrm{c}}+2\left\langle|\nabla f|^{4}\right\rangle_{\mathrm{c}}}{\sigma_{0}^{2} \sigma_{1}^{4}},
	\end{aligned}
	\end{equation}
	where $\left\langle...\right\rangle_{\mathrm{c}}$ denotes the connected part or the cumulant. 
 
\subsection{Numerical evaluation of MFs on \texttt{HEALPix} maps \label{sec:Measure}}
    Several different ways of extracting the MFs from pixelated maps have been suggested in the literature.
	One could for instance count the number of pixel clusters with specific patterns in the map (see, e.g., \cite{Mecke:1996,Mantz_2008,Goring:2013qya}) or identify connected ensembles of pixels by scanning through the map~\cite{Ducout:2012it}. 
	
	In this work, we adopt the method proposed by Schmalzing \& Gorski \cite{Schmalzing:1997uc}: the MFs of a field $f(\bf{x})$ at threshold $\nu$ can be expressed as surface integrals,
	
	\begin{equation}\label{V0_map1}
	   V_{0}(\nu) = \int_{\mathbb{S}^2} \mathrm{d}a \;\Theta(f-\nu),
	\end{equation}
	
	\begin{equation}\label{V1_map1}
	   V_{1}(\nu) =\dfrac{1}{4} \int_{\mathbb{S}^2} \mathrm{d}a \;\left| \nabla f \right| \delta(f-\nu),
	\end{equation}
	
	\begin{equation}\label{V2_map1}
	   V_{2}(\nu) =\dfrac{1}{2\pi} \int_{\mathbb{S}^2} \mathrm{d}a \; \left| \nabla f \right| \delta(f-\nu) \mathcal{K} ,
	\end{equation}		
	
	where $\Theta$ is the Heaviside function and $\mathcal{K}$ the geodesic curvature. The gradient of the field $\left| \nabla f \right|$ and $\mathcal{K}$ can be expressed in terms of derivatives of the field, leading to the following expressions for $V_1$ and $V_2$: 
	
	\begin{equation}\label{V1_map2}
	   V_{1}(\nu) =\dfrac{1}{4} \int \mathrm{d}a \;\delta(f-\nu) \sqrt{f_{\theta}^{2}+f_{\phi}^{2}},
	\end{equation}
	
	\begin{equation}\label{V2_map2}
	   V_{2}(\nu) =\dfrac{1}{2\pi} \int \mathrm{d}a \; \delta(f-\nu) \dfrac{2f_{\theta}f_{\phi}f_{\theta\phi}-f_{\theta}^{2}f_{\phi\phi}-f_{\phi}^{2}f_{\theta\theta}}{f_{\theta}^{2}+f_{\phi}^{2}}.
	\end{equation}	
	We represent the maps using the standard \texttt{HEALPix}~\cite{Gorski:2004by,Zonca2019} hierarchical tessellation of the sphere and compute the derivatives of $f$ using the \texttt{HEALPix} routine \texttt{alm2map\_der1}. In the remainder of this section, we will demonstrate that our numerical methods are consistent with analytical expectations on simulated Gaussian random fields.

\subsection{\textit{Planck}-like Gaussian lensing maps  \label{sec:gaussian}}
    To generate our covariance matrix and demonstrate the validity of our method of evaluating MFs for {\it Planck}-like maps, we use the \texttt{HEALpix} routine \texttt{synfast} to generate full-sky (no mask) maps of Gaussian realisations based on the \Planck FFP10 fiducial model which assumes a base $\Lambda$CDM cosmology with parameter values ($\Omega_\mathrm{b}h^{2} = 0.02216$, $\Omega_\mathrm{cdm} h^{2} = 0.1203$, \mbox{$A_\mathrm{s} = 2.119\times10^{9}$}, \mbox{$n_\mathrm{s} = 0.96$}, $h = 0.67$, $\sum m_{\nu}=0.06~\text{eV}$)\footnote{The sixth free parameter of the base $\Lambda$CDM model, the optical depth to recombination $\tau$, has no impact on gravitational lensing observables.}~\cite{Planck:2018lkk} at a \texttt{HEALpix} resolution of $N_\mathrm{side} = 512$. 
    
    Considering the poor signal to noise at large multipoles, we limit the multipole range of our simulated maps to $8 \leq \ell \leq 400$, consistent with the conservative multipole range used in Ref.~\cite{Planck:2018lbu}. 	After removing the monopole of these maps, we compute their MFs with the method discussed in Section~\ref{sec:Measure} and compare the average of the MFs over all maps with the analytical expectation for the MFs, calculated from the power spectrum using Equations~(\ref{V0_PS})-(\ref{tau}). 
    	
\begin{figure}[ht]
	   \centering
	   \includegraphics[width=1.0\textwidth]{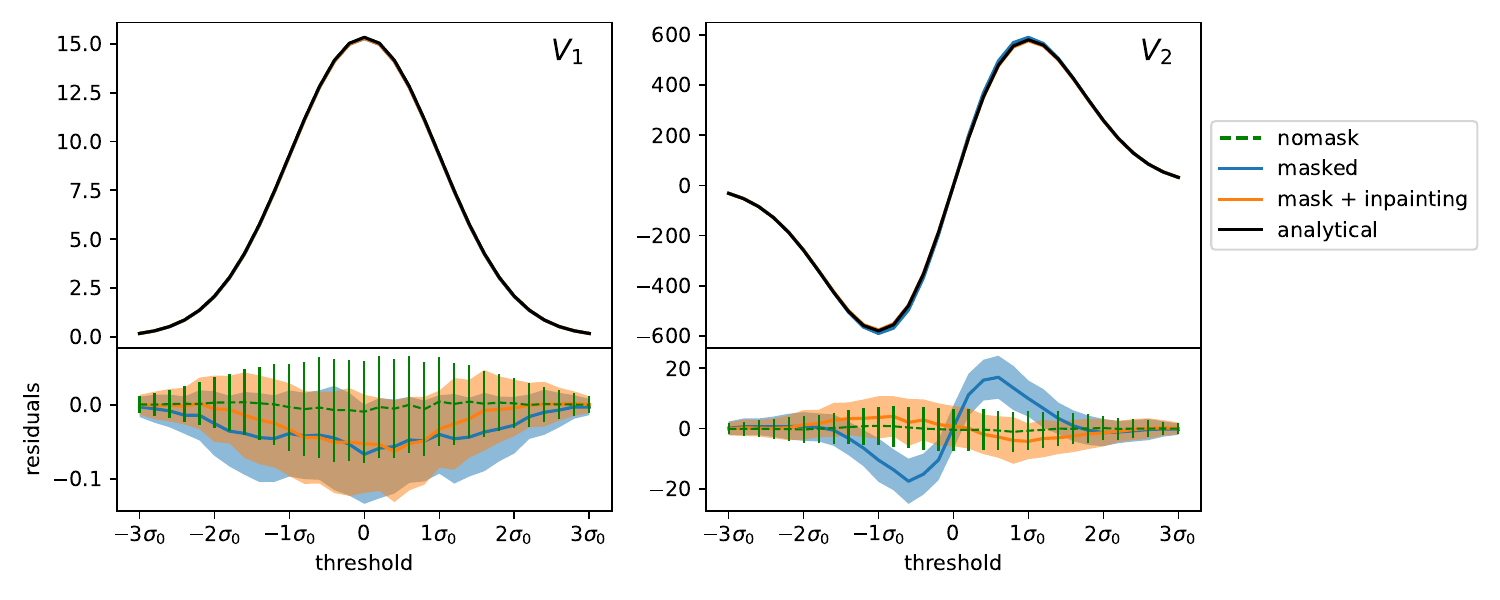}
    	\caption{This plot compares the analytical prediction of $V_1$ and $V_2$ with the average of the MFs measured from 100 simulated maps for three different cases: (i) a full sky simulation without masking, (ii) a full sky simulation masked with the \Planck SMICA mask, and (iii) a full sky simulation masked with the \Planck SMICA mask and inpainted along the boundary of the mask. In the lower panels we show the residuals for each of these cases with respect to the analytical prediction. The coloured bands indicate the standard deviation of the simulated maps' MFs. \label{fig:Gaussian_nomask}}
	\end{figure}

    We show a comparison between MFs measured from simulated maps and the corresponding analytical predictions in Figure~\ref{fig:Gaussian_nomask} as green dashed and black solid lines, respectively.  Note that in this figure we have rescaled the threshold $\nu$ with the theoretical standard deviation $\sigma_0$ calculated from the theoretical power spectrum via Equation~(\ref{sigma}), i.e., $\hat{\nu} = \nu/\sigma_0$.  The excellent consistency between numerical measurements and analytical predictions is evident.

\subsection{Partial sky coverage \label{sec:partial_sky}}
    \begin{figure}[ht]
	    \centering
   	    \begin{subfigure}[b]{0.48\textwidth}
		\centering
		\raisebox{-0.42cm}{\includegraphics[width=\textwidth]{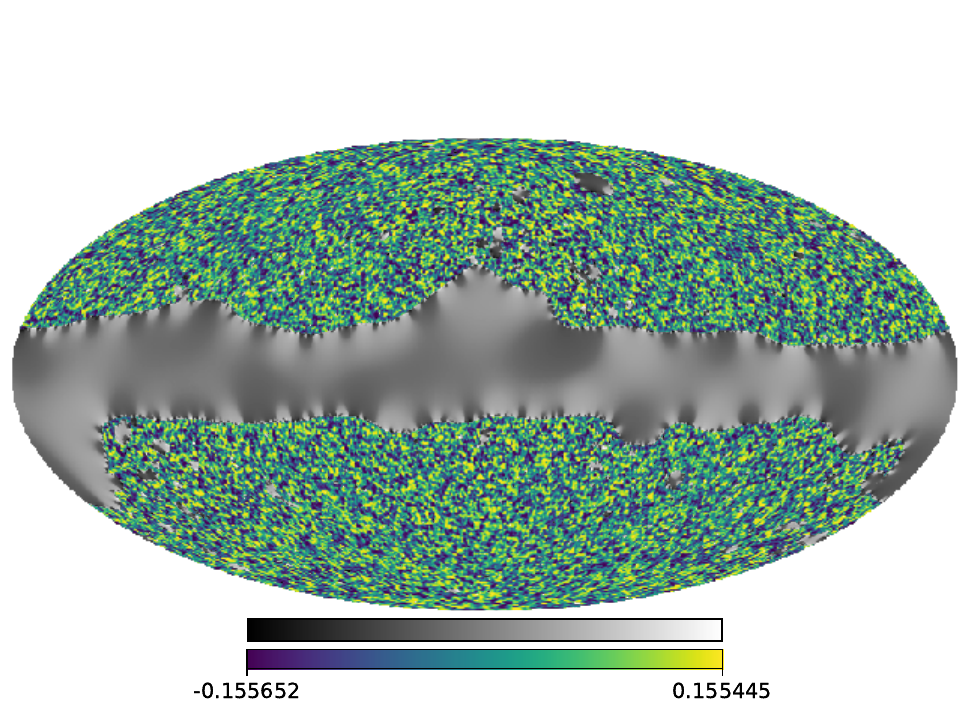}}
		\end{subfigure}
   		\begin{subfigure}[b]{0.48\textwidth}
		\centering
		\raisebox{-0.42cm}{\includegraphics[width=\textwidth]{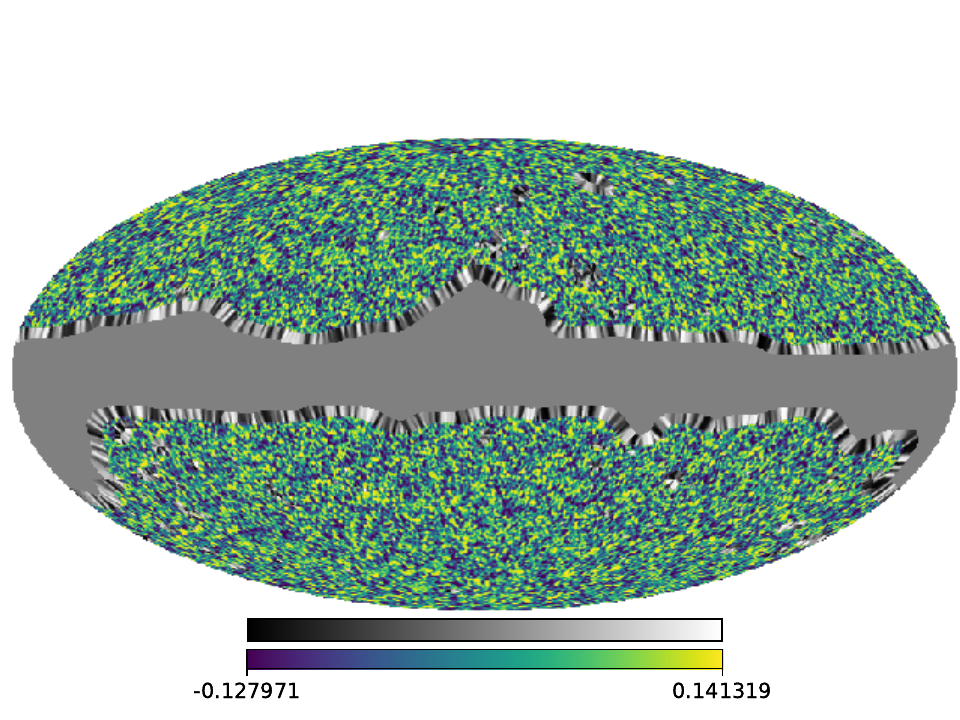}}
		\end{subfigure}
   	    \caption{\textit{Left:} the \Planck 2018 lensing convergence map, including default  inpainting in the masked area. \textit{Right}: a simulated lensing convergence map, inpainted using the algorithm described in the text.  The field in the unmasked region is represented with a blue-green-yellow colour scale; the inpainted field in the masked region uses a grey-scale colour map. \label{fig:inpaint}}
	\end{figure}

    So far, we have taken the field to cover the entire sky. However, in practice one can expect to measure a lensing signal only on part of the sky.  The \Planck lensing map, for instance, covers a sky fraction of $f_\mathrm{sky} = 0.671$, with the final mask being a combination of a SMICA-based confidence mask, a Galactic mask and a point source mask~\cite{Planck:2018lbu}.  In principle, this does not pose a significant obstacle for the prediction and evaluation of MFs, since technically the expressions for $V_0$, $V_1$ and $V_2$ in Equations~\eqref{MF0}-\eqref{MF2} and \eqref{V0_map1}-\eqref{V2_map1} are surface densities -- instead of integrating over the entire sphere, the integration just needs to be performed over the unmasked area.
  
    However, unless proper care is taken, the presence of edges induced by a mask can potentially lead to a serious bias in the numerical evaluation of the map's derivatives, which then propagates to the MFs $V_1$ and $V_2$ as well\footnote{Since $V_0$ is evaluated by simply counting pixels, it is not affected by masking.}, as shown by the blue lines in Figure~\ref{fig:Gaussian_nomask}.         
        
    This problem can be mitigated by inpainting the masked areas before evaluating the derivatives.  
    We apply a simple iterative inpainting procedure to our masked simulated CMB lensing maps: pixels within the masked area with at least four unmasked neighbours are assigned a value equal to the mean of their neighbouring pixels; this step is repeated 50 times, at which point all pixels within $\sim 5$ degrees of the mask boundary are filled.  An example of a thus inpainted simulated map is presented in the right panel of Figure~\ref{fig:inpaint}.\footnote{Note that the \Planck lensing convergence map already comes with the masked areas inpainted, see the left panel of Figure~\ref{fig:inpaint}.}  As demonstrated in Figure~\ref{fig:Gaussian_nomask}, our inpainting method reduces the bias to below the percent level.

\section{Using MFs for cosmological parameter inference \label{sec:likelihood}}

    Having validated our numerical method of evaluating MFs by showing its agreement with analytical expectations, we now proceed to discuss how MFs can be used for the inference of cosmological parameters.  Here, we will be specifically interested in the matter density parameter $\Omega_\mathrm{m}h^{2}$ and the amplitude of the power spectrum of primordial curvature perturbations $A_\mathrm{s}$).  However, before applying our approach to real observations, we will again consider simulated CMB lensing data, as in the previous section.

\subsection{A basic MF likelihood \label{sec:basic_like}}
\begin{figure}[htbp]
        \centering
        \begin{subfigure}{1\textwidth}
            \centering
            \includegraphics[width=\linewidth]{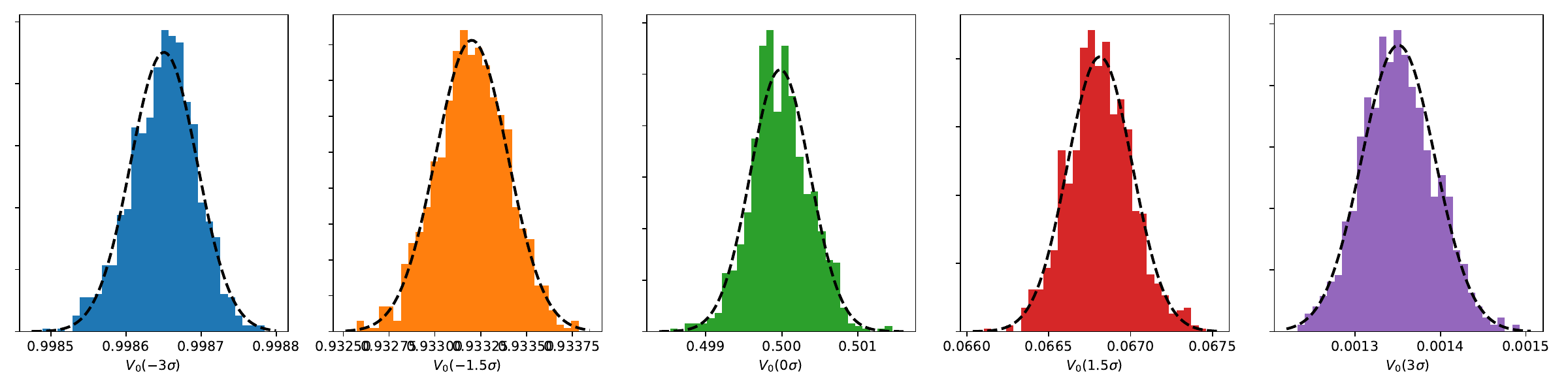}
            \caption{$V_0$}\vspace{10pt}
        \end{subfigure}\hfill
        \begin{subfigure}{1\textwidth}
            \centering
            \includegraphics[width=\linewidth]{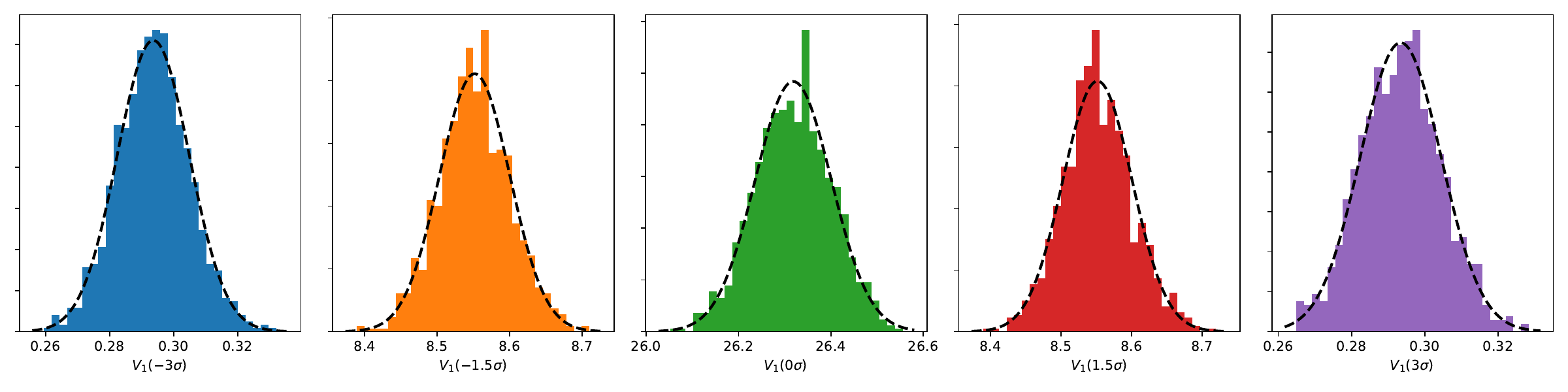}
            \caption{$V_1$}
        \end{subfigure}
        \begin{subfigure}{1\textwidth}
            \centering
            \includegraphics[width=\linewidth]{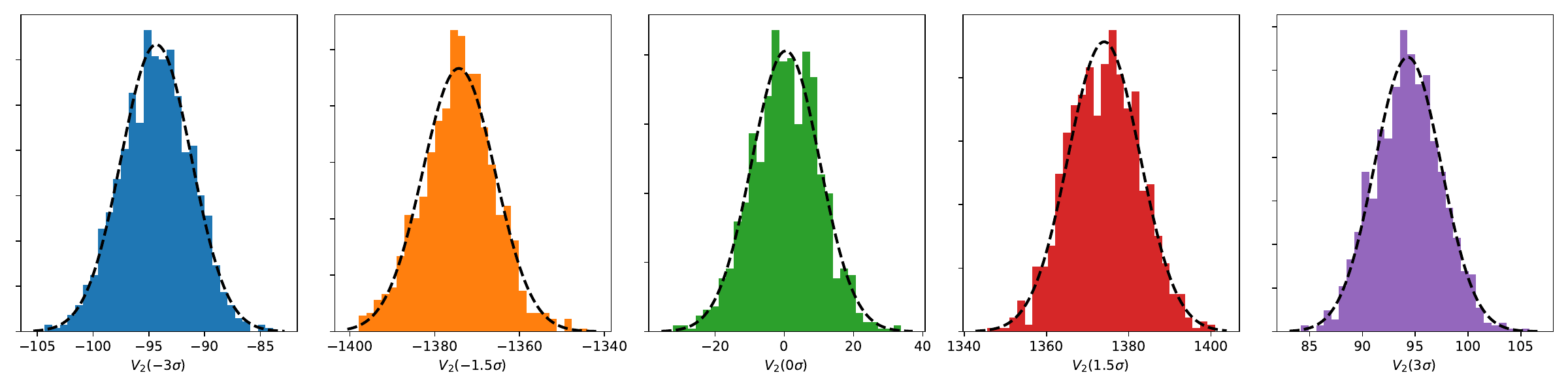}
            \caption{$V_2$}
        \end{subfigure}

        \caption{Histograms and Gaussian fits (dashed black lines) of the MFs at thresholds $\nu \in \left\{-3 \, \sigma_0, -1.5 \, \sigma_0, 0, 1.5 \, \sigma_0, 3\, \sigma_0\right\}$ for 1,000 realisations of full-sky Gaussian random fields, generated from the \textit{Planck} best-fit lensing spectrum.\label{fig:G_likelihood_val}}
    \end{figure}

    First of all, we will need to construct a MF likelihood function. Starting from a data vector $\mathbf{V} = (V_0(\nu_1), ..., V_0(\nu_{n_\mathrm{thr}}), V_1(\nu_1), ..., V_1(\nu_{n_\mathrm{thr}}), V_2(\nu_1), ..., V_2(\nu_{n_\mathrm{thr}}))$ whose entries consist of the three MFs evaluated for a set of $n_\mathrm{thr}$ discrete values of the threshold parameter (we use $n_\mathrm{thr} = 11$ equally-spaced values covering the interval $[-3 \sigma_0, 3 \sigma_0]$; we found that the information gain starts saturating around this number) and noting that over the range of thresholds we are considering, the MFs are reasonably well approximated to follow Gaussian distributions (see Fig.~\ref{fig:G_likelihood_val}), we model the likelihood as
    
    \begin{equation}\label{eq:loglike}
	   -2 \ln \mathcal{L}(\mathbf{V}|\Theta) \propto (\mathbf{V}_{i}-\mathbf{V}_{i}^\mathrm{th}(\Theta)) \, C^{-1}_{ij} \,(\mathbf{V}_{j}-\mathbf{V}_{j}^\mathrm{th}(\Theta)),
	\end{equation}
    where $\mathbf{V}^\mathrm{th}(\Theta)$ is the corresponding theoretical prediction as a function of cosmological parameters $\Theta$.\footnote{In the case of a Gaussian field, these will be calculated via the angular power spectrum calculated with \texttt{CLASS}~\cite{Blas:2011rf} and Equations~\eqref{V0_PS}-\eqref{V2_PS}, \eqref{sigma} and \eqref{tau}.}
    
    Assuming the covariance matrix $C_{ij}$ to be constant over the parameter space of interest, we determine $C_{ij}$ from a set of $N_\mathrm{sim}$ MF vectors $\mathbf{V}^{\text{sim}}$, extracted from simulated maps (in our case a set of 10,000 Gaussian realisations of the fiducial FFP10 cosmology).
    
\begin{figure}[htbp]
        \centering
        \includegraphics[width=0.8\textwidth]{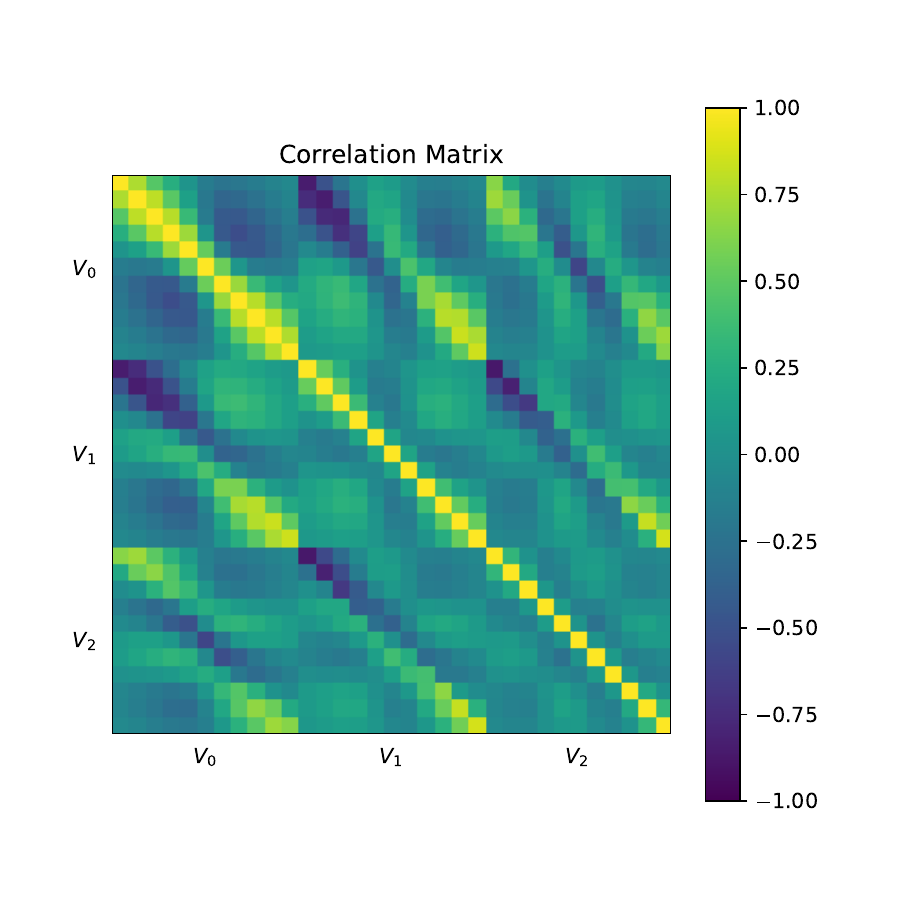}            
        \caption{Correlation matrix for $V_0$, $V_1$ and $V_2$, using $n_\mathrm{thr} = 11$ equally-spaced threshold values covering the interval $[-3 \sigma_0, 3 \sigma_0]$ and estimated from 10,000 Gaussian realisations of {\it Planck}-like lensing maps. \label{fig:corrmatrix_simple}}
    \end{figure}

    Note that for finite $N_\mathrm{sim}$, the na\"ive expression
	\begin{equation}\label{covmatrix}
	\hat{C}_{ij} = \left\langle \left(\mathbf{V}_{i}^{\text{sim}}-\left\langle \mathbf{V}_{i}^{\text{sim}}\right\rangle\right)\left(\mathbf{V}_{j}^{\text{sim}}-\left\langle \mathbf{V}_{j}^{\text{sim}}\right\rangle\right) \right\rangle 
	\end{equation}
    underestimates the true covariance.  This bias can be fixed with the following correction~\cite{Hartlap:2006kj}:
    \begin{equation} \label{eq:covmat_corrected}
    	C_{i j}^{-1} = \frac{N_\mathrm{sim}-p-2}{N_\mathrm{sim}-2} \, \hat{C}_{i j}^{-1} \text{ for } p<N_\mathrm{sim}-2,
    \end{equation} 
    where $p$ is the rank of $C$ (i.e., here, $p = 3 \, n_\mathrm{thr}$).  We show the resulting correlation matrix $c_{ij} = C_{ij} \, (C_{ii} C_{jj})^{-1/2}$ in Figure~\ref{fig:corrmatrix_simple}.

\subsection{A needlet-filtered likelihood\label{sec:needlet}}    
    While the MFs describe the morphological features of a map, they are a priori rather insensitive to scale information.  Nonetheless, this information can become accessible if one decomposes the original map using scale-dependent filters before evaluating the MFs on the individual filtered maps.  Here we opt to decompose the \Planck convergence map using spherical needlets~\cite{Narcowich2006LocalizedTF,Marinucci:2007aj}, following the approach of Ref.~\cite{Planck:2019evm} who applied the same method to a MF analysis of \Planck temperature and polarisation maps.

\begin{figure}[ht]
		\centering
		\includegraphics[width=0.84\textwidth]{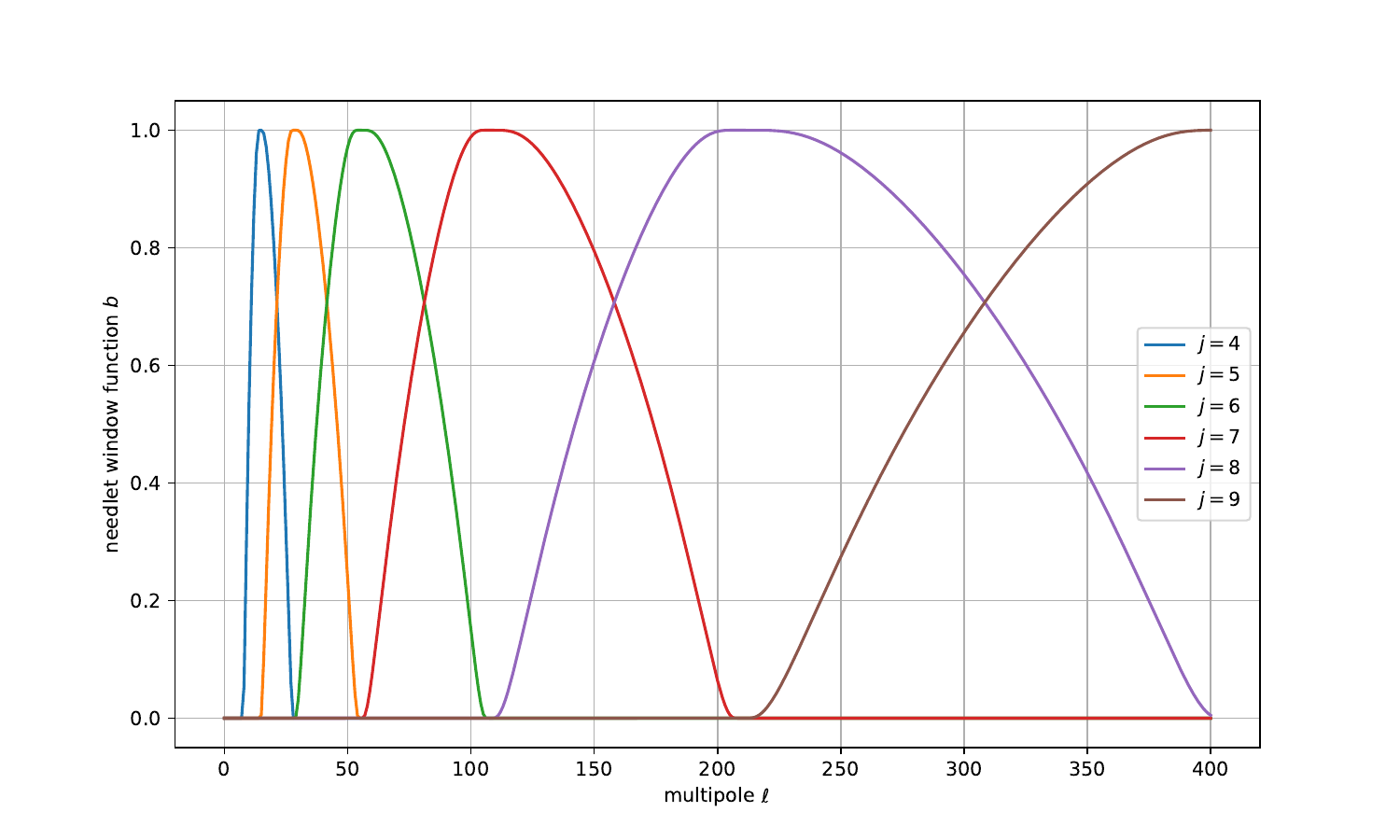}
		\caption{Window functions used in our needlet decomposition, using the form proposed in Ref.~\cite{Marinucci:2007aj} with band width parameter $B = 1.95$ and needlet scale index $j \in \left\{ 4,5,6,7,8,9 \right\}$. \label{fig:needlet}}
	\end{figure}

    Given a map with spherical harmonic expansion coefficients $a_{\ell m}$, needlet-filtered maps can be constructed via a convolution of the map with a window function $b$ in harmonic space~\cite{Marinucci:2007aj}:
    \begin{equation} \label{eq:needletcomponents}
	   \beta_{j}(\hat{\boldsymbol{n}})=\sum_{\ell=B^{j-1}}^{B^{j+1}} b\left(\frac{\ell}{B^{j}}\right) \sum_{m = -\ell}^{\ell} a_{\ell m} Y_{\ell m}(\hat{\boldsymbol{n}}),
	\end{equation}
	where $Y_{\ell m}$ are the associated Legendre polynomials, $j \in \mathbb{N}$ is the scale index, $B$ is the band width parameter of the needlet and the window function $b$ is smooth with support on $\ell \in \left[ B^{j-1}, B^{j+1} \right]$ and satisfying $\sum_{j} b^{2}\left(\frac{\ell}{B^{j}}\right)=1$ for all $\ell$.   We follow the window function construction method suggested in Ref.~\cite{Marinucci:2007aj} with $B = 400^{1/9} \simeq 1.95$.  For $n_j = 6$ filtered maps with $j \in \left\{ 4,5,6,7,8,9 \right\}$ this choice covers the entire multipole range of interest, as shown in Figure~\ref{fig:needlet}.  

\begin{figure}[ht]
		\centering
		\includegraphics[width=0.84\textwidth]{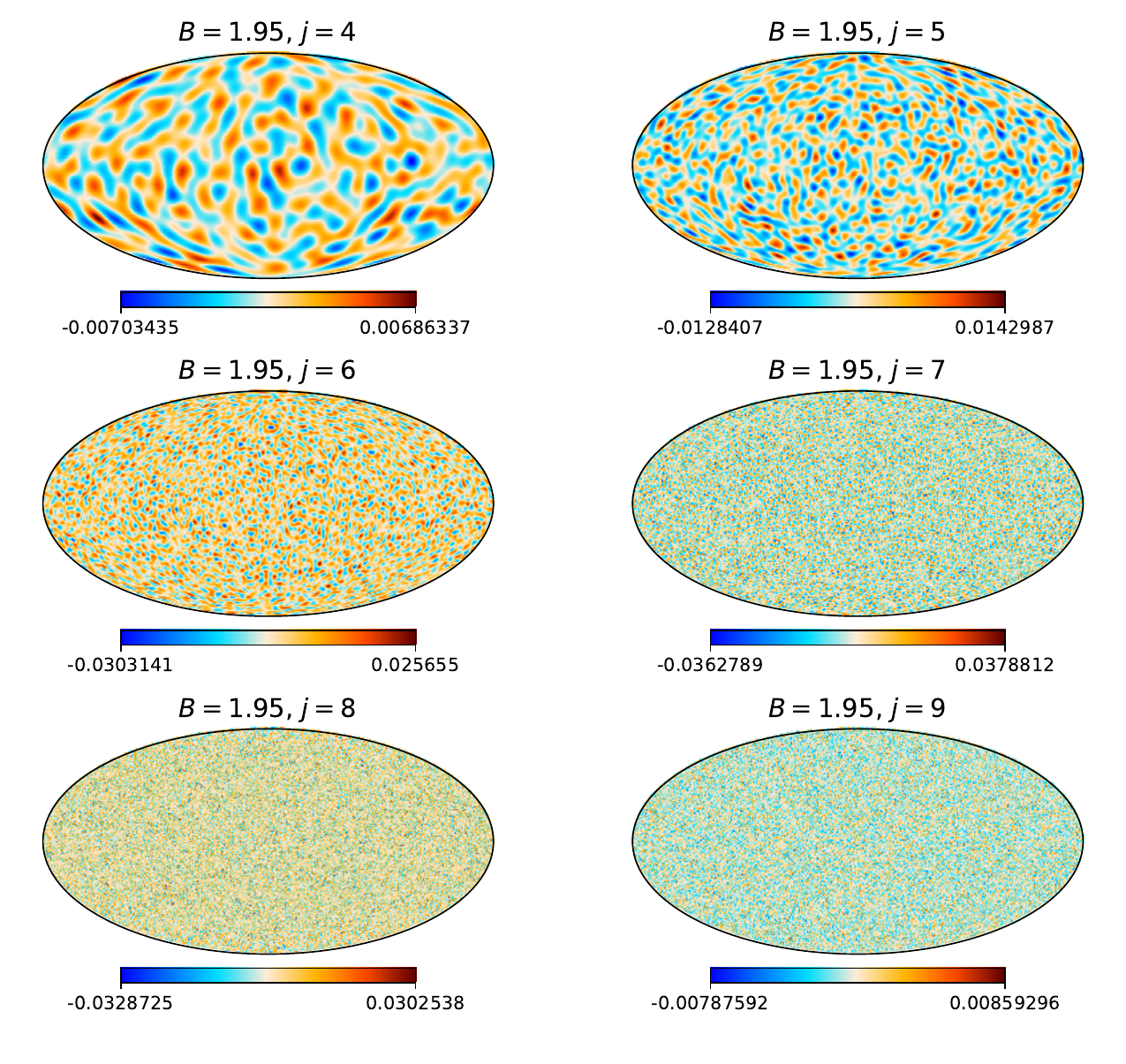}
		\caption{Needlet filtered simulated Gaussian CMB lensing map with $B = 400^{1/9} \simeq 1.95$, $j \in \left\{ 4,5,6,7,8,9 \right\}$. \label{fig:needlet_map}}
    \end{figure}

    For the numerical implementation of the needlet decomposition we employ the Python package \texttt{MTNeedlet}~\cite{Duque:2019jxt}\footnote{\texttt{https://javicarron.github.io/mtneedlet/}} which itself makes use of the \texttt{HEALpix} routine \texttt{almxfl} to perform the filtering.  The resulting maps are illustrated in Figure~\ref{fig:needlet_map} – it can be clearly seen that the larger the needlet index, the finer the structure of the corresponding filtered map.

    The likelihood for $n_j$ needlet-decomposed maps can be defined analogously to the basic one discussed in Section~\ref{sec:basic_like}, except that the vector $\mathbf{V}$ now contains $3 n_j n_\mathrm{thr}$ instead of $3 n_\mathrm{thr}$ entries. We checked that Gaussianity is still a good approximation in this case as well (see also Refs.~\cite{Cammarota:2013pya,SHEVCHENKO2023268}, which derive Central Limit Theorems for excursion sets of needlet-transformed fields in the high-frequency ($j \rightarrow \infty$) limit).   

\begin{figure}[htbp]
        \centering            
        \includegraphics[width=0.84\textwidth]{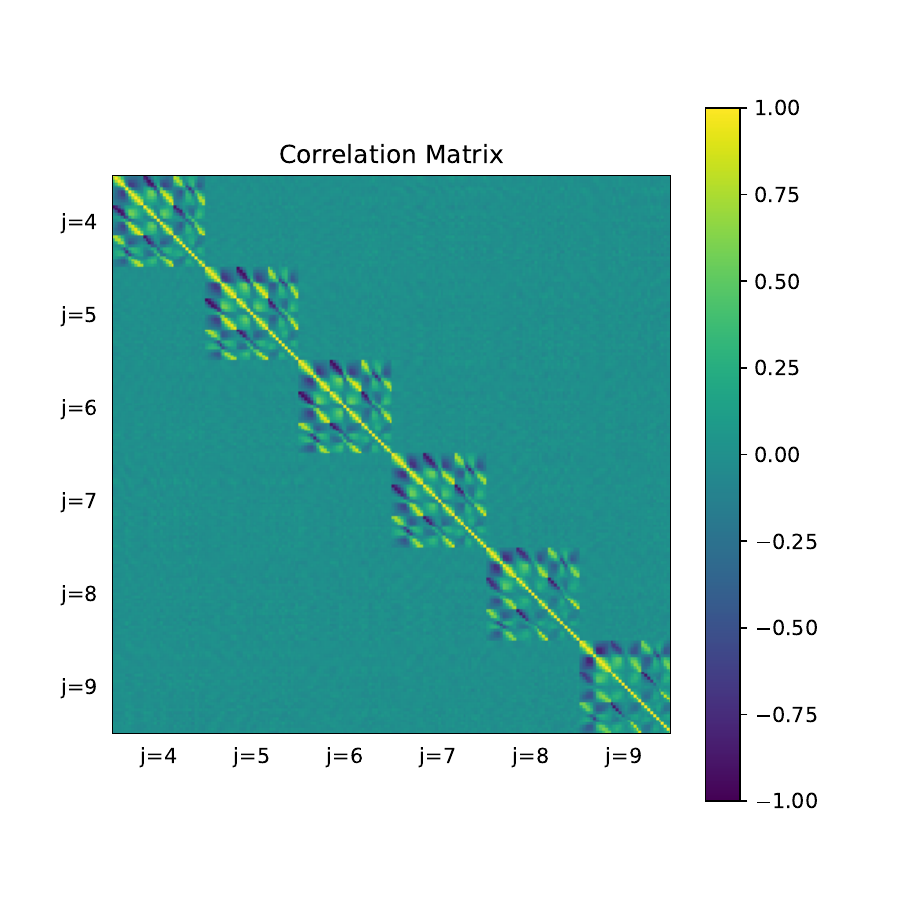}
        \caption{Correlation matrix for $V_0$, $V_1$ and $V_2$ over six needlet-filtered maps, using $n_\mathrm{thr} = 11$ equally-spaced threshold values covering the interval $[-3 \sigma_0, 3 \sigma_0]$ and estimated from 10,000 Gaussian realisations of {\it Planck}-like lensing maps.  The additional covariance due to a correction for non-Gaussianity of the noise (see the discussion in Section~\ref{sec:nG_val} below) is also included. \label{fig:corrmatrix_needlet}}
    \end{figure}

    Note that for full-sky maps, there is no overlap between the window functions of needlets with $\Delta j > 1$, and therefore one does not expect them to be correlated.  In the case of incomplete sky coverage with a realistic CMB mask, the correlations remain very weak~\cite{Marinucci:2007aj}.  We nonetheless compute the full covariance matrix, based on 10,000 realisations of simulated Gaussian lensing maps (see Figure~\ref{fig:corrmatrix_needlet}).

    \subsubsection{Accessing scale information: an example}
\begin{figure}[htbp]
        \centering
        \begin{subfigure}{0.6\textwidth}
            \centering
            \includegraphics[width=\linewidth]{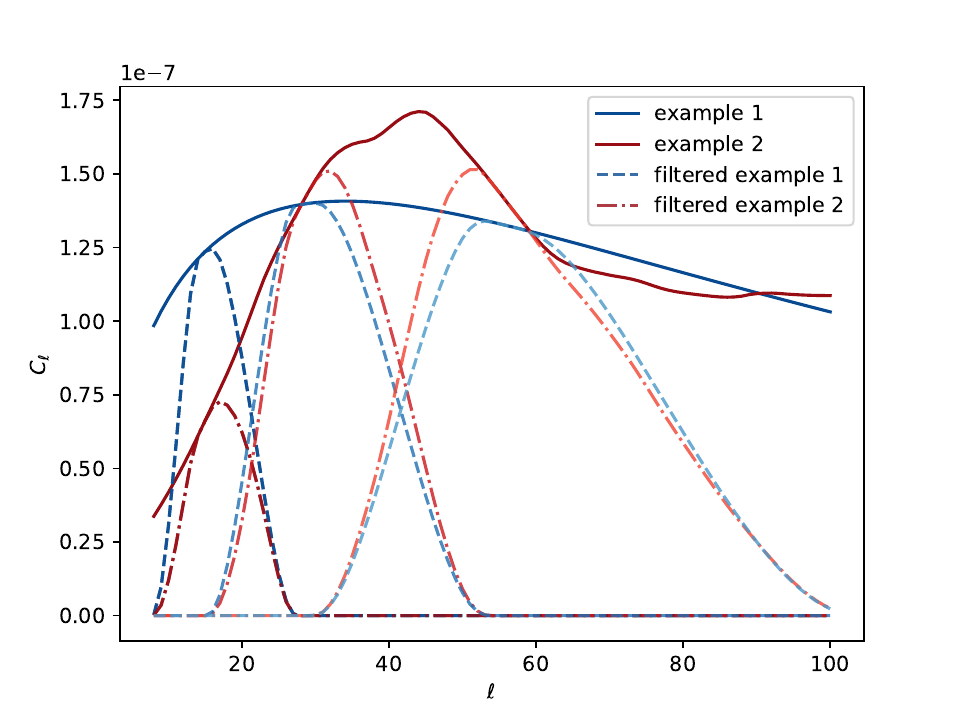}
            \caption{Two examples of power spectrum that with the same $\sigma_0$ and $\sigma_1$, plotted over the interval $8 \leq \ell \leq 100$. The corresponding power spectra of the filtered maps with needlet parameters $B=1.95$, $j \in \left\{ 4,5,6 \right\}$ are shown as dashed and dot-dashed lines.}

            \vspace{10pt}
            \label{fig:Cl_same_sigma}
        \end{subfigure}\hfill
        \begin{subfigure}{0.75\textwidth}
            \centering
            \includegraphics[width=\linewidth]{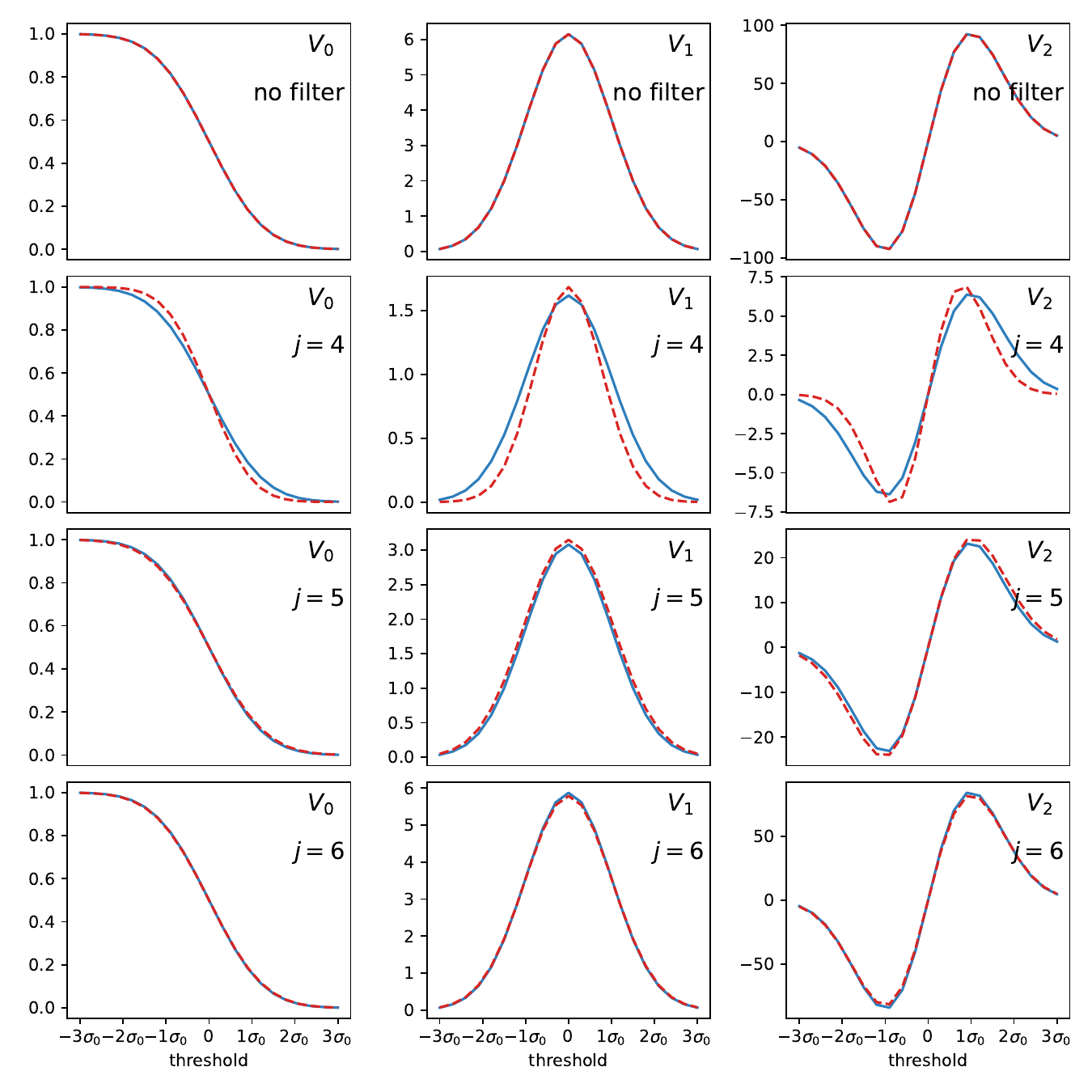}
            \caption{The MFs for the above power spectra. \label{fig:MF_same_sigma}}

        \end{subfigure}

        \caption{Power spectra and MFs for two fields that share identical $\sigma_0$ and $\sigma_1$.}\label{fig:same_sigma0_sigma1}
    \end{figure}

    Let us demonstrate the added discriminatory power due to the filtering procedure with an illustrative example. As is evident from Equations~(\ref{V0_PS})-(\ref{V2_PS}), for a Gaussian random field the MFs are completely determined by $\sigma_0$ and $\sigma_1$. MFs are therefore unable to distinguish between two fields with the same $\sigma_0$ and $\sigma_1$, even if their power spectra may be very different.  This is for instance the case for the two power spectra given by the red and blue solid lines in Figure~\ref{fig:Cl_same_sigma}: the top row of Figure~\ref{fig:MF_same_sigma} shows that the MFs of the unfiltered maps coincide exactly.
    
    After filtering with spherical needlets with $B=1.95$ and $j \in \left\{ 4,5,6 \right\}$, the power spectra of the filtered maps can be calculated via 
    \begin{equation}
    C_\ell^{(j)} = b^2\left(\frac{\ell}{B^{j}}\right) C_\ell,
    \end{equation}
    and their respective $\sigma_0$ and $\sigma_1$ are given by
    \begin{equation}\label{sigma_nd}
    \sigma_0^2(j)=\sum_{B^{j-1} \leq \ell \leq B^{j+1}} b^2\left(\frac{\ell}{B^j}\right) \frac{2 \ell+1}{4 \pi} C_{\ell}, 
    \end{equation}
    \begin{equation}\label{sigma1_nd}
    \sigma_1^2(j)=\sum_{B^{j-1} \leq \ell \leq B^{j+1}} b^2\left(\frac{\ell}{B^j}\right) \frac{2 \ell+1}{4 \pi} \ell(\ell+1) C_{\ell} .
    \end{equation}
    The resulting MFs of the filtered maps (see Figure~\ref{fig:MF_same_sigma}) are not identical anymore, with the difference being most pronounced for the $j=4$ needlets here.  Given sufficiently high-quality data, the filtered maps can thus break the degeneracy present in the unfiltered case.

\subsection{Parameter inference from Gaussian random fields}\label{section:Val_G}
    With the likelihood functions of Sections~\ref{sec:basic_like} and \ref{sec:needlet} at hand, we can now discuss how to use them to constrain cosmological parameters.  For now we will take the lensing map to obey Gaussian statistics; this assumption will be dropped in the next Section.

    As a means of validating our method, we generate sets of synthetic MF data based on the \Planck FFP10 fiducial model plus \Planck noise by averaging over the MFs extracted from 1,000 Gaussian realisations of masked (unfiltered and needlet-filtered, respectively) lensing maps.  Likewise, we construct the corresponding covariance matrices from 10,000 simulations. The priors we adopt for the cosmological parameters are same as in the \Planck lensing-only likelihood analysis~\cite{Planck:2018lbu}, shown in Table~\ref{table:cosmplogy_prior}.  We infer the posterior probabilities of the cosmological parameters via standard Markov Chain Monte Carlo sampling, as implemented in the \texttt{Cobaya} package~\cite{2019ascl.soft10019T,Torrado_2021}.

    \begin{table}[h!]
        \centering
        \begin{tabular}{c c c}
            \hline
            \textbf{Parameter} & \textbf{Prior range} & \textbf{Prior type} \\ \hline
            $\Omega_\mathrm{b} h^2$ & $0.0222 \pm 0.0005$ & \multirow{2}{*}{Gaussian}\\
            $n_\mathrm{s}$ & $0.96 \pm 0.02$  \\ 
            \hline
            $\Omega_\mathrm{c} h^2$ & $[0.001,0.7]$ &  \multirow{3}{*}{top hat} \\ 
            $100 \,\theta_\mathrm{MC}$ & $[0.5,10]$  \\ 
            $\ln(10^{10} A_\mathrm{s})$ & $[1.61,3.91]$  \\ 
            \hline
            $\sum m_{\nu}$ & $0.06$ & fixed \\ \hline
        \end{tabular}
        \caption{Cosmological parameters and their associated prior probabilities.  In addition, we impose lower and upper limits on the Hubble parameter $H_0 \in [40, 100]$ (note that $H_0$ is a derived parameter in this analysis and not directly varied).  \label{table:cosmplogy_prior}}
    \end{table}

\begin{figure}[htbp]
        \centering
        \begin{subfigure}{.48\textwidth}
            \centering
            \includegraphics[width=\linewidth]{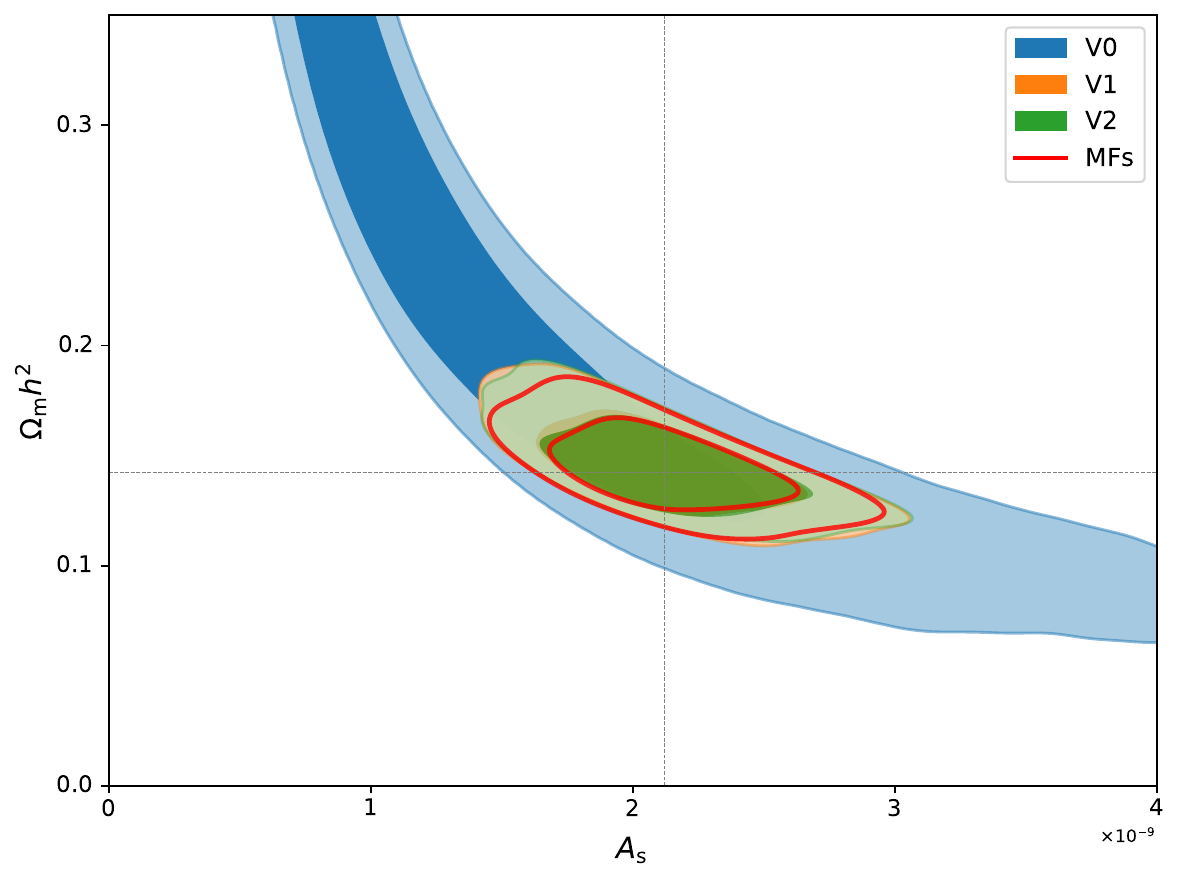}
            \caption{Constraints for individual MFs and joint MFs with the basic likelihood function.}\vspace{10pt}
            \label{fig:gauss_basic}
        \end{subfigure}\hfill
        \begin{subfigure}{.48\textwidth}
            \centering
            \includegraphics[width=\linewidth]{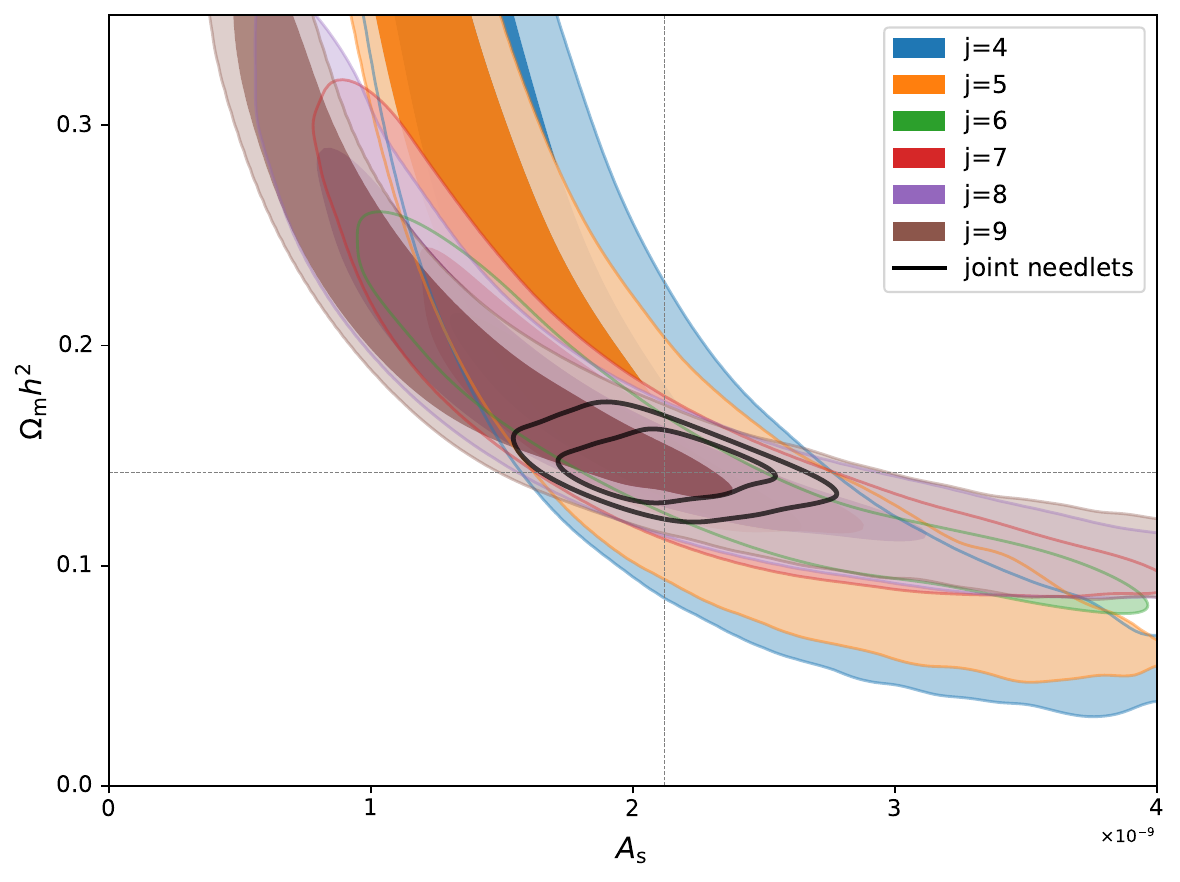}
            \caption{Constraints for joint MFs with individual needlet-filtered maps and the combination of all six of them.}
            \label{fig:gauss_needlet}
        \end{subfigure}
        \caption{Marginalised posterior probability isocontours for $A_\mathrm{s}$ and $\Omega_\mathrm{m}h^2$ with \textit{Planck}-like simulated Gaussian MF data. Fiducial parameter values are indicated by dashed grey lines, and the inner and outer contours delineate 68\%- and 95\%-credible regions, respectively. \label{fig:gauss_val}}
    \end{figure}

    The resulting joint constraints for the basic MF likelihood in the $(A_\mathrm{s}, \Omega_\mathrm{m}h^2)$-plane are shown in Figure~\ref{fig:gauss_basic}, broken down into the contributions of the individual MFs (blue, orange and green filled contours), as well as the constraint obtained from the combination of all three MFs (red outline contours).  It is not surprising that $V_0$ has substantially weaker constraining power than $V_1$ and $V_2$, since in the Gaussian case, $V_0$ is only sensitive to the field's standard deviation $\sigma_0$ (see Equation~\eqref{V0_PS}), so it cannot distinguish between parameter combinations that leave $\sigma_0$ invariant; in our case, this combination is approximately $\omega_\mathrm{m} A_\mathrm{s}^{0.8}$. 
     
    The other two MFs, $V_1$ and $V_2$, are sensitive to the cosmological parameters via $\sigma_0$ and the ratio $\sigma_1/\sigma_0$ (see Equations~\eqref{V1_PS} and \eqref{V2_PS}) and are thus able to break this degeneracy.  However, with both MFs being sensitive to the same combinations of $\sigma_0$ and $\sigma_1$, they do not provide independent information and lead to practically identical constraints\footnote{This property is specific to the Gaussian case due to the non-Gaussian corrections to $V_1$ and $V_2$ being subject to different combinations of skewness and kurtosis parameters.} – this also explains why the combination of all MFs is only very slightly more constraining than $V_1$ or $V_2$ by themselves.

    For the needlet likelihood, we show the corresponding results in Figure~\ref{fig:gauss_needlet}, with the coloured contours denoting constraints from individual needlet maps (and all three MFs), and the black outline contours the constraints from the full joint likelihood.  The plot nicely illustrates that the degeneracy direction depends on the needlet index: with larger $j$, corresponding to smaller scales being probed, the degeneracy direction becomes less steep.  So even though each individual needlet map has little constraining power, the combination of all of them can effectively break the individual degeneracies and leads to a visible improvement over the unfiltered basic likelihood.

    For both likelihoods, Figure~\ref{fig:gauss_val} shows that we retrieve an unbiased estimate of the input parameter values.

\subsection{Parameter inference from maps with Gaussian signal and weakly non-Gaussian noise \label{sec:nG_val}}
    When considering realistic lensing maps, the assumption of Gaussianity is an overly simplistic one since one would expects the signal itself to be mildly non-Gaussian, and the same applies to the reconstruction noise.  In case of the \Planck data, it is the latter that makes the dominant contribution the non-Gaussianity of the reconstructed map~\cite{Planck:2018lbu}.  This is somewhat unfortunate because in principle the non-Gaussian component of the signal contains cosmological information too, but here we will need to regard the non-Gaussianity as a nuisance whose contribution to the MFs must be removed to ensure unbiased parameter estimates.
        
    As in the previous Section, we will first validate our approach on simulated data, considering the case of a Gaussian signal with non-Gaussian noise. We construct non-Gaussian convergence maps $\kappa_\mathrm{nG}$ from Gaussian ones $\kappa_\mathrm{G}$ via the following pixel-by-pixel transformation:
    \begin{equation}
      	\kappa_\mathrm{nG} = \kappa_\mathrm{G} + f_{\text{nl}}\left(\kappa_\mathrm{G}^2  -\left<\kappa_\mathrm{G}^2\right>\right) + g_{\text{nl}} \, \kappa_\mathrm{G}^3.
    \end{equation} 
    The non-linearity parameters are set to $f_{\text{nl}}=1$ and $g_{\text{nl}}=3$; this setting reproduces skewness and kurtosis parameters of roughly the same magnitude as those of the \Planck map.  Our synthetic non-Gaussian MF data are then obtained by averaging over the MFs of 1,000 non-Gaussian map realisations.

\begin{figure}[htbp]
        \centering
        \includegraphics[width=0.8\textwidth]{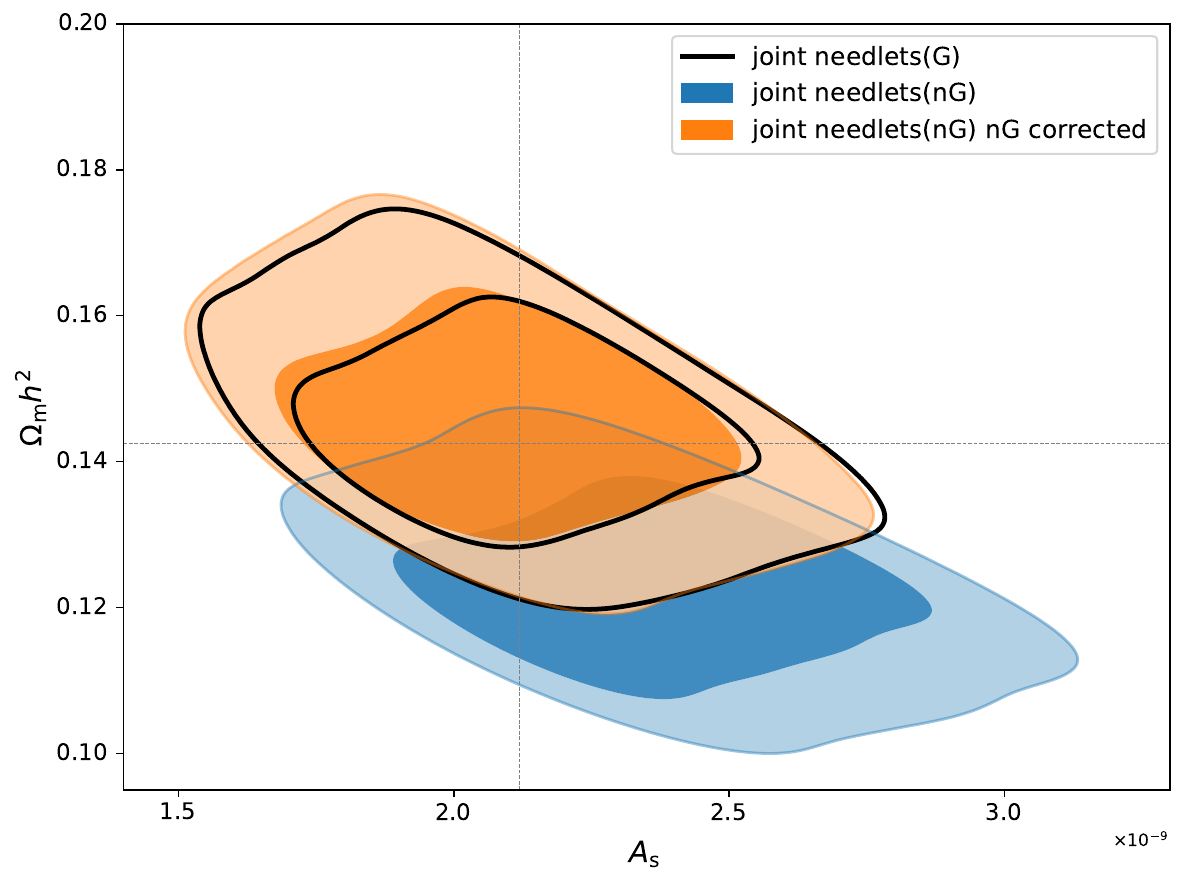}
            
        \caption{Marginalised posterior probability isocontours for $A_\mathrm{s}$ and $\Omega_\mathrm{m}h^2$ with \textit{Planck}-like simulated non-Gaussian MF data with (orange) and without (blue) correction – we also plot the corresponding result for Gaussian data for comparison (black outlines). The full needlet-filtered likelihood is used here, but only the orange contours contain the additional contribution to the covariance matrix from the non-Gaussian correction.  Fiducial parameter values are indicated by dashed grey lines, and the inner and outer contours delineate 68\%- and 95\%-credible regions, respectively.  \label{fig:ng_val}}
    \end{figure}

    If one were to na{\"i}vely fit these data using a purely Gaussian model, the resulting parameter estimates pick up a bias of $\mathcal{O}(1)$ standard deviation, as can be seen from the blue contours in Figure~\ref{fig:ng_val}. The solution to this problem lies in adding the expected non-Gaussian contribution to the MFs, $\widehat{\delta \mathbf{V}}_\mathrm{nG}$, to the Gaussian theoretical prediction $\mathbf{V}^\mathrm{th}_\mathrm{G}(\Theta)$:
    \begin{equation}
        \mathbf{V}^\mathrm{th}_\mathrm{nG}(\Theta) = \mathbf{V}^\mathrm{th}_\mathrm{G}(\Theta) + \widehat{\delta \mathbf{V}}_\mathrm{nG},
    \end{equation}
    or, equivalently, subtracting it from the MF data.  For realistic lensing data, $\mathbf{V}^\mathrm{th}_\mathrm{G}(\Theta)$ will not be analytically tractable, but can be estimated from simulations instead.  In case of the \Planck data, one can use the lensing maps that form part of the Full Focal Plane (FFP10) suite of simulations~\cite{Planck:2018lkk}; in this Section, we will use a set of $N_\mathrm{sim} = 300$ (identical to the number of available FFP10 maps, see discussion below) non-Gaussian maps for this purpose, averaging over the difference between the MFs of the maps, $\mathbf{V}^i_\mathrm{nG}$, and the corresponding Gaussian prediction for the MFs of each map, $\mathbf{V}^i_\mathrm{G}$ (calculated via the expressions given in Section~\ref{sec:grf} and using the actual power spectrum of each map): 
    \begin{equation}
        \widehat{\delta \mathbf{V}}_\mathrm{nG} \simeq \frac{1}{N_\mathrm{sim}} \, \sum_i^\mathrm{N_\mathrm{sim}} \left( \mathbf{V}^i_\mathrm{nG} - \mathbf{V}^i_\mathrm{G} \right).
    \end{equation}
    Given that the non-Gaussian parameters of individual realisations of maps are random variates, subtracting the mean will introduce an additional uncertainty, which can be quantified in terms of a covariance matrix.  Assuming the variance due to the correction to be uncorrelated with the statistical variance in the Gaussian case (Equation~\eqref{eq:covmat_corrected}), we can co-add the two covariance matrices in the expression for the likelihood (Equation~\eqref{eq:loglike}).  As can be gleaned from the orange contours in Figure~\ref{fig:ng_val}, our correction removes the bias and does not lead to a notable increase in the overall uncertainty on $A_\mathrm{s}$ and $\Omega_\mathrm{m}h^2$.

\section{Constraints with the \Planck CMB lensing map\label{sec:results}}

\subsection{Properties of the \Planck lensing map's non-Gaussianity \label{sec:Planck_noise}}
\begin{figure}[htbp]
		\centering
		\begin{subfigure}[b]{0.75\textwidth}
			\centering
			\includegraphics[width=\textwidth]{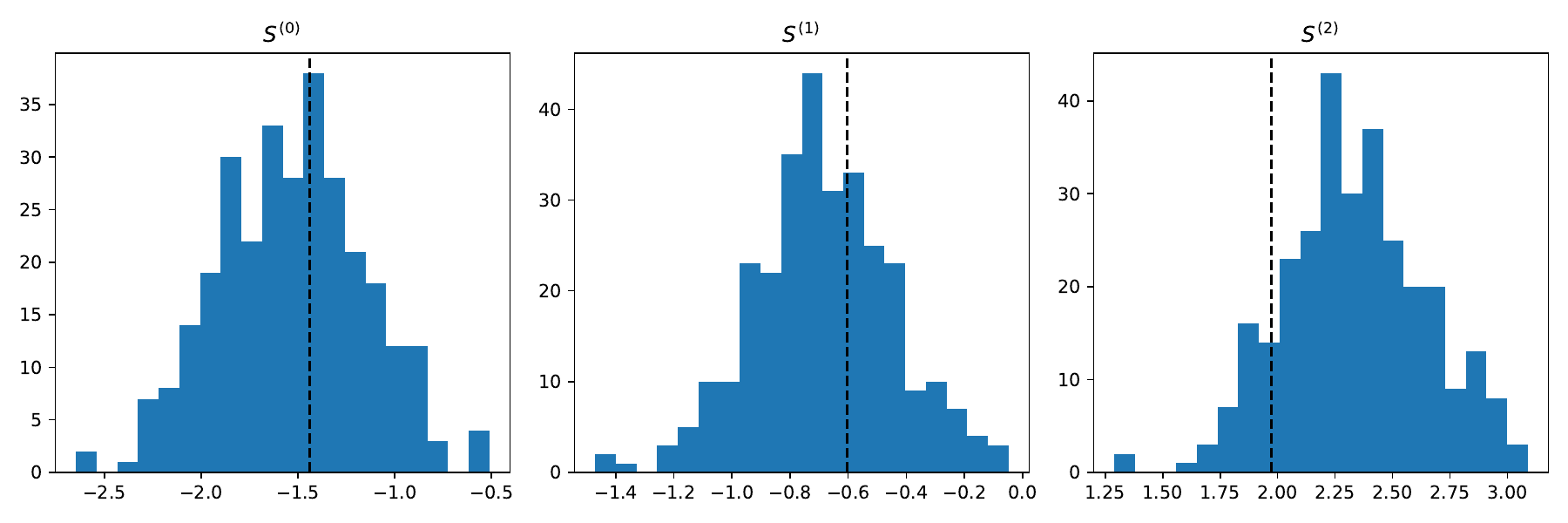}
		\end{subfigure}
		\begin{subfigure}[b]{1\textwidth}
			\centering
			\includegraphics[width=\textwidth]{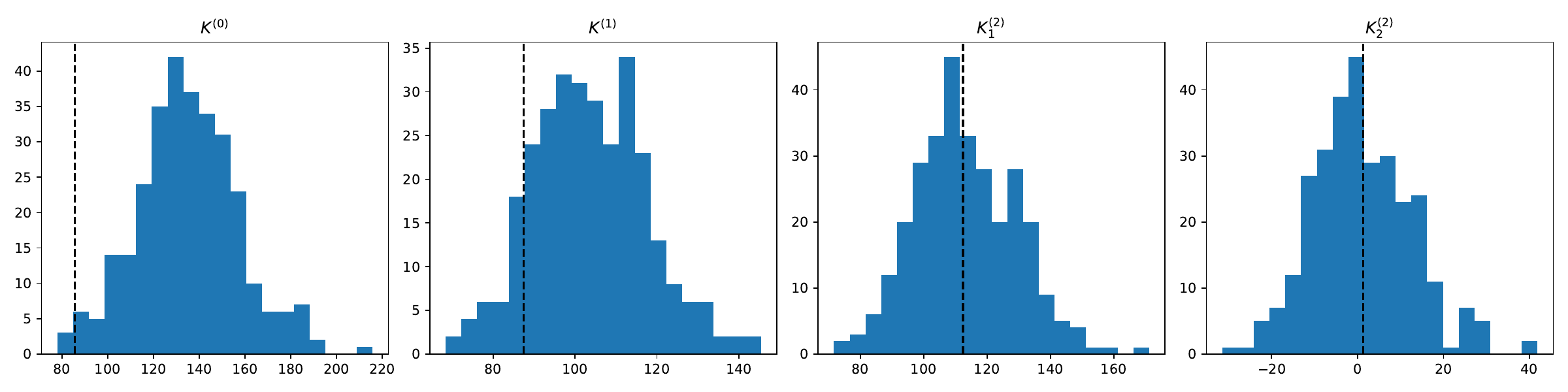}
		\end{subfigure}
    	\caption{Histograms of skewness and kurtosis parameters of the FFP10 lensing maps.  The vertical dashed lines are the corresponding values for the actual \Planck CMB lensing map.  \label{fig:FFP_ng}}
	\end{figure}

    Before we apply our MF-based analysis pipeline to the \Planck lensing map, let us first briefly discuss some of its non-Gaussian properties.  In the context of this discussion, we will interpret a map's skewness and kurtosis parameters as a measure of their non-Gaussianity, noting that for a purely Gaussian map, their expectation values would be zero.  This is a statement about ensemble averages though, and as long as we only consider a single realisation of a full-sky map, even an underlying Gaussian field will present with non-zero skewness and kurtosis parameters.  However, their sample variance can be determined from simulations, and for the \Planck map, we have the set of 300 FFP10 simulations at our disposal.  We show the histograms of the non-Gaussianity paramters for the FFP10 lensing maps in Figure~\ref{fig:FFP_ng}, along with their actual values in the \Planck map.  
    
    \noindent 
    These results imply:
    \begin{itemize}
        \item{The \Planck lensing map is statistically significantly non-Gaussian.}
        \item{The FFP10 lensing maps are also statistically significantly non-Gaussian.  Since in the simulations the signal is assumed to be Gaussian, their non-Gaussianity must have resulted from reconstruction noise.}
        \item{The \Planck map's values of skewness and kurtosis parameters are consistent with the distributions observed in the FFP10 simulations.}
        \item{The \Planck map is consistent with a Gaussian signal and its non-Gaussianity is due to reconstruction noise.}
    \end{itemize}
    This means our correction method from Section~\ref{sec:nG_val} can safely be applied to the \Planck map.
    
\begin{figure}[htbp]
		\centering
        {\includegraphics[width=0.5\textwidth]{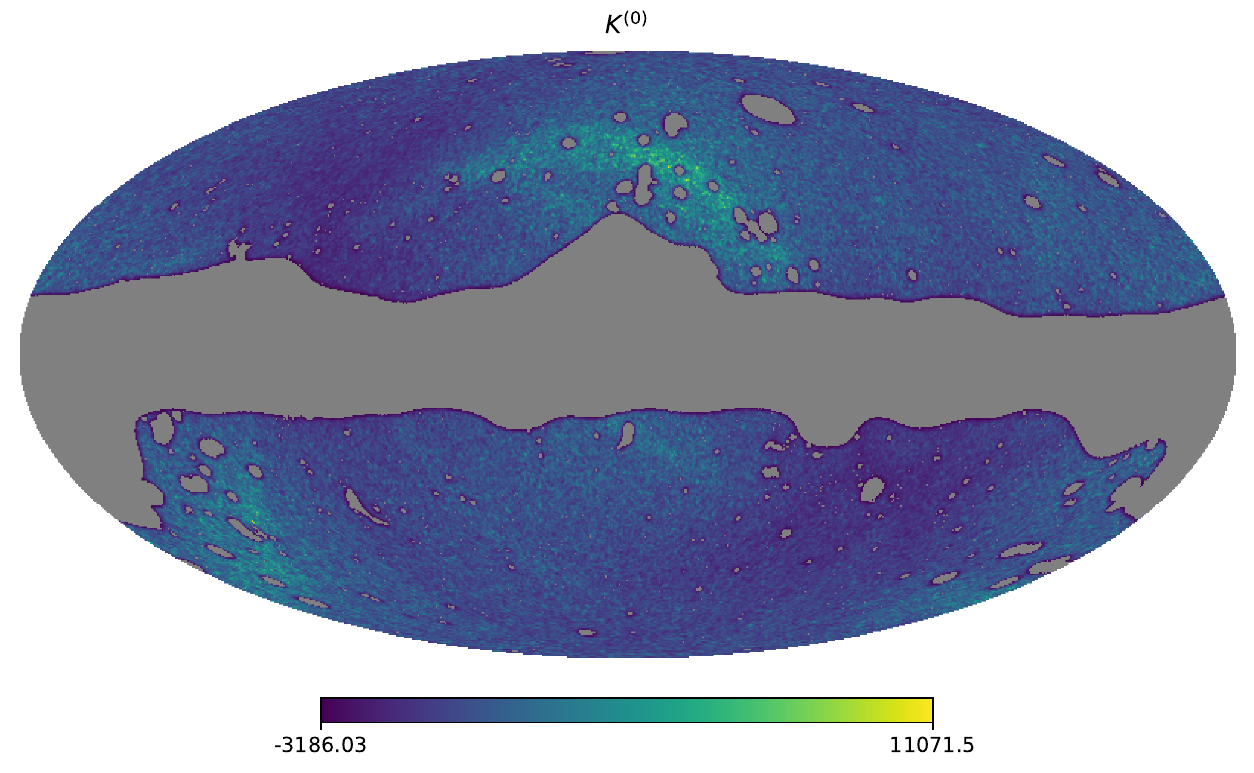}}
	   \caption{Map of $K^{(0)}_\mathrm{local}$, averaged over the 300 FFP10 lensing simulations. \label{fig:FFP_anisotropy}}
	\end{figure}

    Another way of looking at this issue is via the presence (or absence) of spatial correlations between noise and non-Gaussianity.  Owing to the nature of \textit{Planck}'s scanning strategy, the \Planck temperature and polarisation maps are subject to anisotropic noise ((illustrated for the example of the noise of the \Planck temperature map in the upper panel of Figure~2 in Ref.~\cite{Planck:2018lbu}), and therefore the lensing reconstruction noise will inherit this property as well.\footnote{Regarding the statistical anisotropy of the data, instead of averaging the MFs over the entire CMB sky, one could in principle consider further refining the analysis by evaluating the MFs on small patches which are each approximately statistically isotropic – similar in spirit to the approach of Ref.~\cite{Novaes:2016wyx}.  However, unless one neglects the correlations between different patches, a likelihood analysis would be numerically prohibitive.}  If the non-Gaussianity of the maps were of cosmological origin due to lensing, one would expect on average all directions in the sky to contribute equally to the overall non-Gaussianity.  As an example, we can consider for instance a ``local'' definition of the first kurtosis parameter $K^{(0)}_\text{local}=\left(\kappa^{4}-3\sigma_{0}^{4}\right)/\sigma_{0}^{6}$, and averaging over the 300~FFP10 simulations, it is clear from Figure~\ref{fig:FFP_anisotropy} that $K^{(0)}_\text{local}$ does not obey statistical isotropy and a comparison with Figure~2 of Ref.~\cite{Planck:2018lbu} shows that it is highly correlated with the noise of the temperature map.  The other skewness and kurtosis parameters show a similar behavior.
    
\subsection{The \Planck map's MFs and cosmological parameter constraints with the MF likelihood}
\begin{figure}[htbp]
        \centering
        \includegraphics[width=0.99\linewidth]{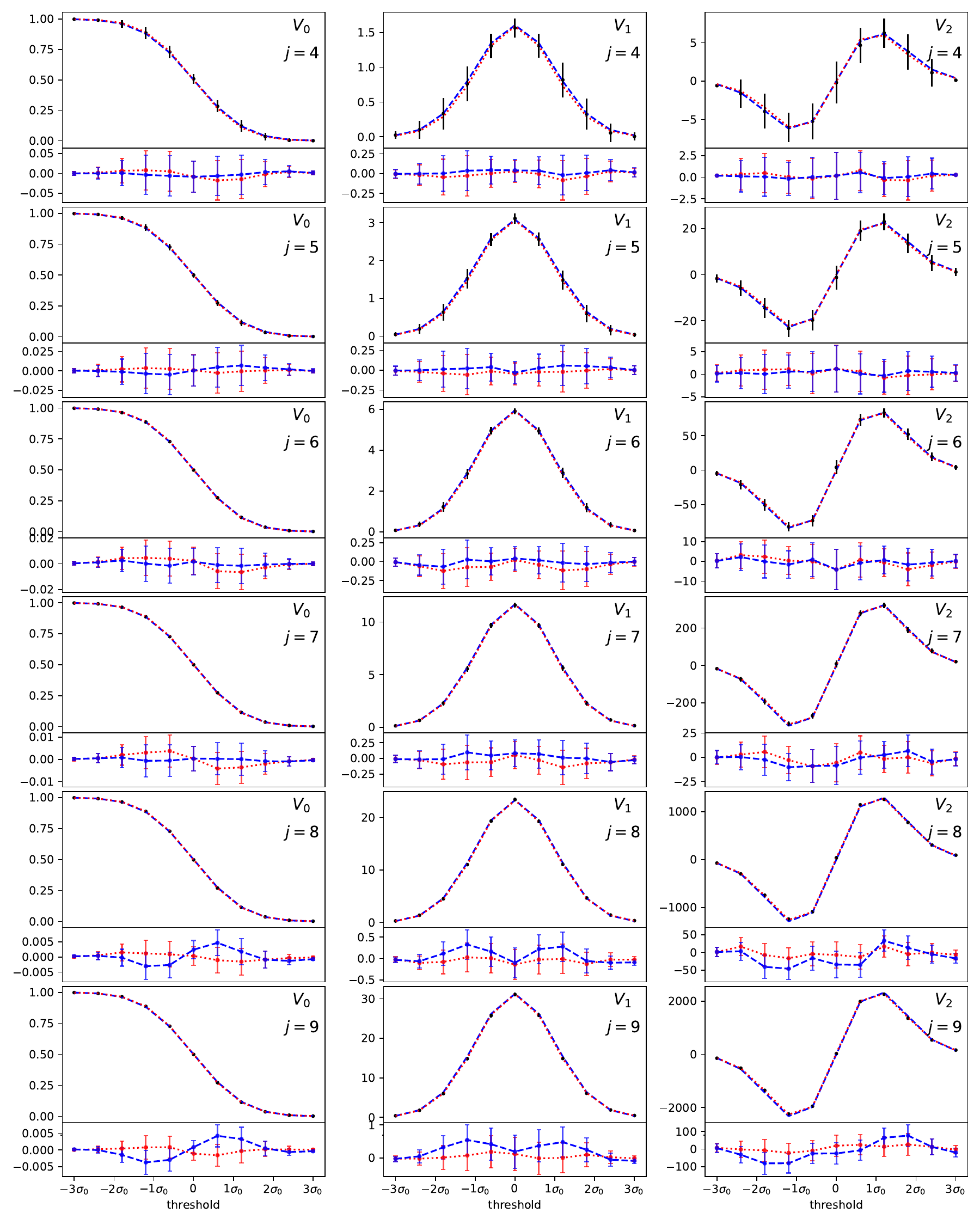}
        \caption{MFs of the \Planck lensing map after applying needlet filters with needlet scale indices $j \in \left\{ 4,5,6,7,8,9 \right\}$ and the corresponding residuals with respect to the analytical predictions for the best-fit cosmological parameters (dotted red lines include the correction for non-Gaussian noise, dashed blue lines do not).  Data points are plotted with $5\sigma$~error bars.  \label{fig:Planck_MFs}}
    \end{figure}

    The MFs for our six needlet-filtered versions of the \Planck convergence map are shown in Figure~\ref{fig:Planck_MFs}.  Notably, the smaller-scale needlet maps ($j = 7,8,9$) display both a smaller uncertainty and a more significant non-Gaussian correction than the larger-scale maps ($j = 4,5,6$).    
    While the signal-to-noise of the unbinned power spectrum peaks around $\ell \sim 40$ (covered by the $j=5$ and $j=6$ needlet maps) and increases at smaller scales, this is offset by the higher needlet indices covering wider multipole ranges (see Figure~\ref{fig:needlet}).
    
    It is also important to keep in mind that these data points are not independent; there are quite substantial correlations for points with neighbouring threshold values, particularly for $V_0$ (as shown in Figure~\ref{fig:corrmatrix_simple}).  Additionally, with the MFs' response to changes in cosmological parameters being non-linear, apparent sub-percent-level accuracy in MF data points will not necessarily translate to similarly tight constraints on cosmological parameters.
   
    Figure~\ref{fig:Planck_MFs} shows the resulting constraints on $\Omega_\mathrm{m}h^2$ and $A_\mathrm{s}$.  By visual inspection, the posterior contours obtained from the six individual needlet-filtered likelihoods appear compatible with each other and with the constraints from a joint analysis as well.  At a more quantitative level, this is confirmed by the best-fit to the full likelihood, which has an effective $\chi$-squared of $\chi^2_\mathrm{eff} \equiv -2 \ln \mathcal{L} \simeq 209.0$, i.e., a reasonable fit to the data given $n_\mathrm{dof} = 196 + 2 - 5 = 193$\footnote{$3 \, n_j n_\mathrm{thr}$ MF data points plus two Gaussian priors versus five free parameters.} effective degrees of freedom.

\begin{figure}[htbp]
        \centering
  	    \begin{subfigure}[b]{0.48\textwidth}
		  \centering
		  \includegraphics[width=\textwidth]{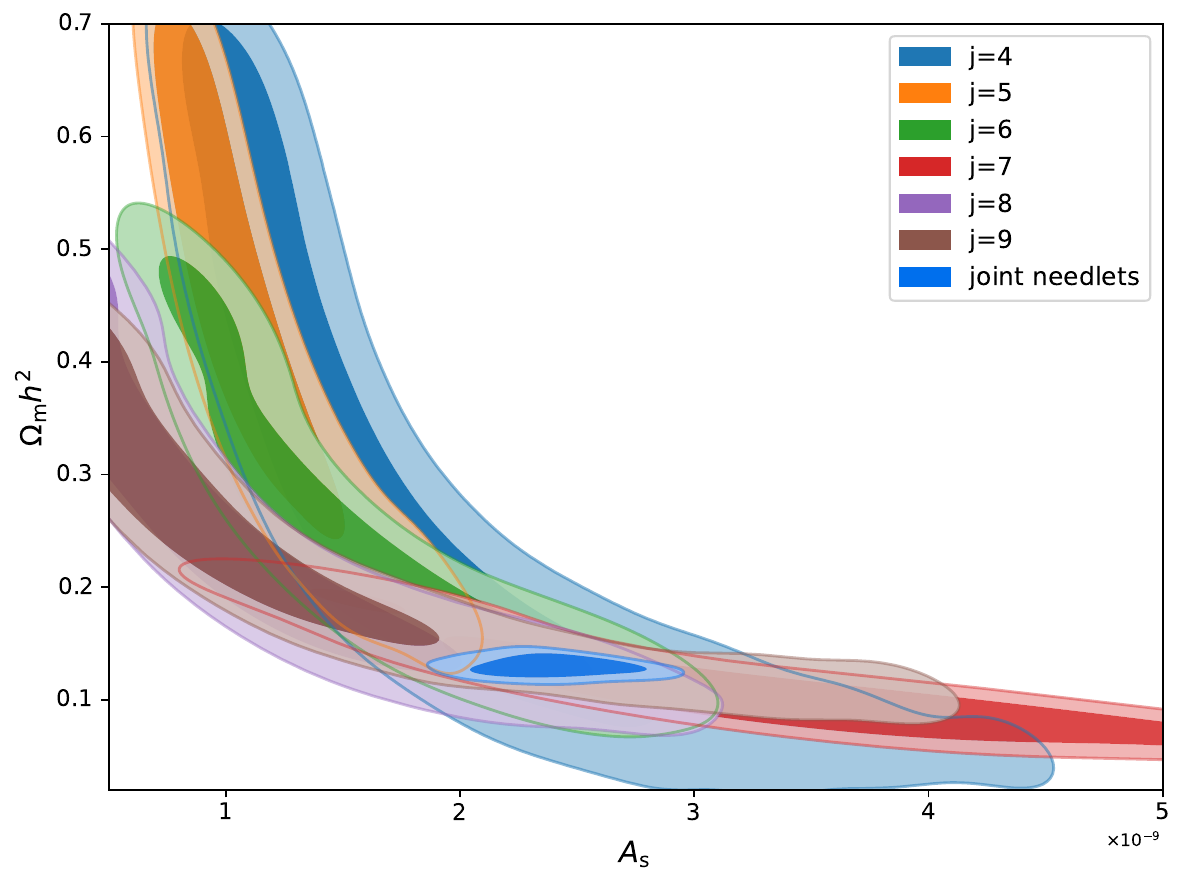}
	    \end{subfigure}
        \begin{subfigure}[b]{0.48\textwidth}
		   \centering
	       \includegraphics[width=\textwidth]{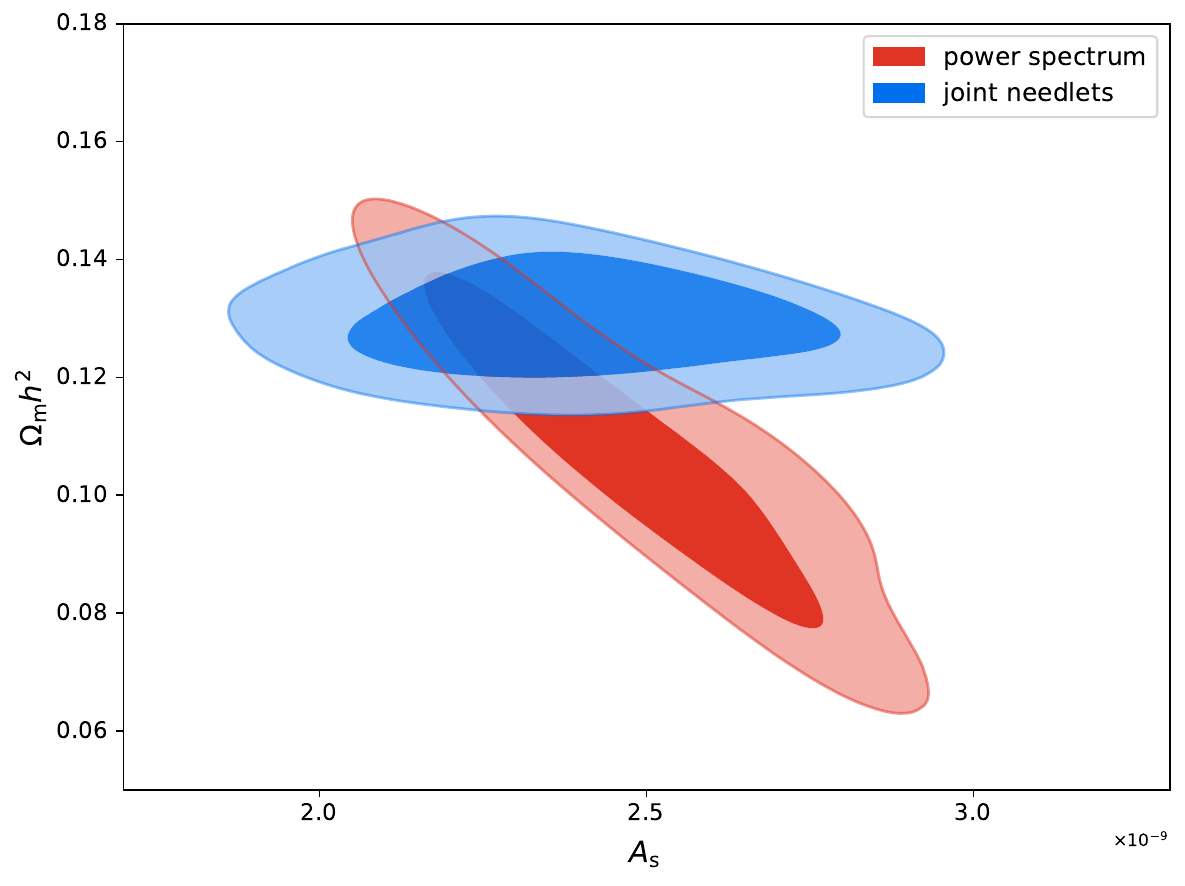}
	    \end{subfigure}
        \caption{Marginalised posterior probability isocontours for $A_\mathrm{s}$ and $\Omega_\mathrm{m}h^2$ given \Planck lensing data.  \textit{Left:} Constraints using the MF likelihood for individual needlet-filtered versions of the \Planck map, as well as from the full likelihood (blue contours).   \textit{Right:} Blue contours are the same as in the left panel, red contours correspond to constraints from the official \Planck power spectrum-based lensing likelihood~\cite{Planck:2018lbu}.  \label{fig:Planck_constraint_wavelet}}
    \end{figure}

    A comparison with constraints from the standard power spectrum-based \Planck lensing likelihood (right panel of Figure~\ref{fig:Planck_constraint_wavelet}) also shows an excellent compatibility of the results, and demonstrates that a MF-based approach can achieve a very similar constraining power, even without tapping into the non-Gaussian cosmological information of a map.

\section{Conclusions \label{sec:conclusions}}
    
    In this paper we have constructed and validated a Minkowski functional-based likelihood approach to the inference of parameters from cosmological observations.  At this stage, it is applicable to maps with Gaussian signal and Gaussian or mildly non-Gaussian noise.  We have demonstrated that the \Planck lensing convergence map falls into this category and after applying our approach to it, find results that are not just in good agreement with those from a power spectrum analysis, but also competitive in terms of sensitivity.
    
    The main appeal of an MF-based analysis lies in its potential generalisation to cases with explicitly non-Gaussian signal, where it will be able to access information in the higher order correlations of the field that is inaccessible to a more standard approach based exclusively on the power spectrum.  
    
    Realising the full potential of the MFs will firstly require observations where the non-Gaussianity of the signal can be distinguished from non-Gaussianity of the noise (which may be true for lensing maps of future CMB surveys, e.g., CMB-S4~\cite{Abazajian:2019eic} or of course other observables such as galaxy weak lensing or density fields).  Secondly, it will require a way to reliably calculate the theoretical prediction of the full non-Gaussian observables as a function of cosmological parameters in a reasonable amount of time.  For CMB weak lensing, that might take the form of numerical full-sky CMB lensing simulations~\cite{Carbone:2007yy, Takahashi:2017hjr}, combined with an emulator, or machine-learning techniques.  We will leave an exploration of these possibilities to future work.

\acknowledgments
    Some of the results in this paper have been derived using the \texttt{healpy} and \texttt{HEALPix} packages.  This research includes computations using the computational cluster \textit{Katana} supported by Research Technology Services at UNSW Sydney~\cite{Katana}.  We thank Steen Hannestad, Bin Hu and Bj\"{o}rn Malte Sch\"{a}fer for fruitful discussions.

\bibliographystyle{apsrev}
\bibliography{MF_journal}
\end{document}